\documentclass[useAMS,usenatbib]{mn2e} \usepackage{amsmath,amssymb}
\usepackage{graphicx,mathptm} \usepackage{times} \usepackage{natbib}
\usepackage{epsfig}
\usepackage[dvips]{color}

\definecolor{Black}{named}{Black}
\definecolor{Red}{named}{Red}

\def\alt{\raise0.3ex\hbox{$\;<$\kern-0.75em\raise-1.1ex\hbox{$\sim\;$}}}
\def\agt{\raise0.3ex\hbox{$\;>$\kern-0.75em\raise-1.1ex\hbox{$\sim\;$}}}

\title[A cross-correlation study of the Fermi-LAT $\gamma$-ray diffuse extragalactic signal]
{A cross-correlation study of the Fermi-LAT $\gamma$-ray diffuse extragalactic signal}

\author[]  {Jun-Qing Xia$^1$, Alessandro Cuoco$^2$,
Enzo Branchini$^{3,4,5}$, Mattia Fornasa$^{6,7}$, Matteo Viel$^{8,9}$ \\
$^1$ Scuola Internazionale di Studi Avanzati,  Via Bonomea 265, I-34136, Trieste, Italy\\
$^2$ Stockholm University - Oskar Klein Center AlbaNova University
Center, Fysikum,
SE-10691, Stockholm, Sweden \\
$^3$ Dipartimento di Fisica, Universit\`a degli Studi ``Roma
Tre'', via della Vasca Navale 84, I-00146 Roma, Italy\\
$^4$ INAF, Osservatorio Astronomico di Brera, Milano, Italy\\
$^5$ INFN, Sezione di Roma Tre, via della Vasca Navale 84, I-00146 Roma, Italy\\
$^6$ Istituto de Astrof\'isica de Andaluc\'ia (CSIC), E18008, Granada, Spain\\
$^7$ MultiDark fellow \\
$^8$ INAF Osservatorio Astronomico di Trieste, Via G. B. Tiepolo 11,
I-34141, Trieste, Italy\\
$^9$ INFN, Sezione di Trieste, via Valerio 2, I-34127, Trieste,
Italy
}

\begin{document}

\maketitle

\label{firstpage}

%%%%%%%%%%%%%%%%%%%%%%%%%%%%%%%%%%%%%%%%%%%%%%%%%%%%%%%%%%%%%%%%%%%%%%%%%%%%%%%%%%%%%%%%%%%%%%%%%%%%%%%%%%%%
%%%%%%%%%%%%%%%%%%%%%%%%%%%%%%%%%%%%%%%%%%%%%%%%%%%%%%%%%%%%%%%%%%%%%%%%%%%%%%%%%%%%%%%%%%%%%%%%%%%%%%%%%%%%
%%%%%%%%%%%%%%%%%%%%%%%%%%%%%%%%%%%%%%%%%%%%%%%%%%%%%%%%%%%%%%%%%%%%%%%%%%%%%%%%%%%%%%%%%%%%%%%%%%%%%%%%%%%%

\begin{abstract}

In this work, starting from 21 months of data from the Fermi-Large
Area Telescope (LAT),  we derive maps of the residual isotropic
$\gamma$-ray emission, a relevant fraction of which is expected to
be contributed  by the extragalactic diffuse $\gamma$-ray background
(EGB). We search for the auto-correlation signals in the above
$\gamma$-ray  maps and for the cross-correlation signal with the
angular distribution of different classes of objects that trace the
large scale structure of the Universe. We compute the angular
two-point auto-correlation function of the residual Fermi-LAT maps
at energies  $E>1$~GeV, $E>3$~GeV and $E>30$~GeV well above the
Galactic plane and find no significant correlation signal. This is,
indeed, what is expected if the EGB were contributed by BL Lacertae
(BLLacs), Flat Spectrum Radio Quasars (FSRQs) or star-forming
galaxies, since, in this case, the predicted signal is very weak.
Then, we search for the Integrated Sachs-Wolfe (ISW) signature by
cross-correlating the Fermi-LAT maps with the WMAP7-Cosmic Microwave
Background map. We find a cross-correlation consistent with zero,
even though the expected signal is larger than  that of the EGB
auto-correlation.  Finally, in an attempt to constrain the nature of
the $\gamma$-ray background  we cross-correlate the Fermi-LAT maps
with the angular distributions of objects that may contribute to the
EGB: QSOs in the SDSS -DR6 (Sloan Digital Sky Survey- Data Release
6) catalog, NRAO VLA Sky Survey (NVSS) galaxies, Two Micron All Sky
Survey  (2MASS) galaxies and Luminous Red Galaxies (LRGs) in the
SDSS catalog. The cross-correlation is always consistent with zero,
in agreement with theoretical expectations, but we find (with low
statistical significance) some interesting features that  may
indicate that some specific classes of objects contribute to the
EGB. A  $\chi^2$ analysis confirms that the correlation properties
of the 21-month data do not provide strong constraints of the EGB
origin. However, the results suggest that the situation will
significantly improve with the 5- and 10-year Fermi-LAT data. The
future EGB analysis will then allow placing significant constraints
on the nature of the EGB and might provide in addition a detection
of the ISW signal.
\end{abstract}

\begin{keywords}
cosmology: theory -- cosmology: observations -- cosmology: large
scale structure of the universe -- gamma rays: diffuse backgrounds
\end{keywords}

%%%%%%%%%%%%%%%%%%%%%%%%%%%%%%%%%%%%%%%%%%%%%%%%%%%%%%%%%%%%%%%%%%%%%%%%%%%%%%%%%%%%%%%%%%%%%%%%%%%%%%%%%%%%

\section{Introduction}

One of the most interesting open problems in astrophysics is the
nature of the diffuse $\gamma$-ray background at GeV energies. The
presence of a diffuse signal was first detected by the OSO-3
satellite \citep{1972ApJ...177..341K} with indications of both
Galactic and isotropic diffuse emissions.  SAS-2
\citep{1973ApJ...186L..99F} and later EGRET
\citep{1998ApJ...494..523S} revealed more clearly the isotropic
component, which is commonly known as extragalactic $\gamma$-ray
background (EGB). Although several local processes have been
proposed to explain this background
\citep{2004JCAP...04..006K,2006ApJ...652L..65M,2007ApJ...664L.143M,2009ApJ...692L..54M},
the EGB is generally believed to be the superposition of
contributions from unresolved extragalactic sources and, perhaps,
diffuse GeV emission processes like the annihilation of dark matter
(DM) particles \citep{2002PhRvD..66l3502U}.  We will, however, in
the following use the abbreviation ``EGB'', even if the
extragalactic origin of this component is still not fully clear.

Blazars are the most numerous population detected by EGRET and the
Fermi Large Area Telescope (Fermi-LAT)
\citep{1999ApJS..123...79H,2009ApJ...700..597A}. They have long been
regarded as the most likely candidates to provide the bulk of the
EGB emission. However, the recent analysis of the First Fermi-LAT
AGN catalog obtained after $\sim1$ year of data taking revealed that
blazars can contribute $23\pm 5$\%(stat.) $\pm$12\%(syst.) of the
EGB in the range between 0.1 and 100 GeV
\citep{2010ApJ...720..435A},  disfavoring the interpretation of the
EGB as mainly consisting of unresolved blazars. DM candidates in
supersymmetric theories (as well as other WIMP candidates) can
annihilate into GeV photons and contribute to the EGB
\citep{2002PhRvD..66l3502U,
2007PhRvD..75f3519A,2009arXiv0908.0195P,Zavala:2009zr,Zavala:2011tt}.
However, the amplitude and shape of the observed EGB spectrum
\citep{2010JCAP...04..014A, 2010PhRvL.104j1101A} together with the
available multi-wavelength and multi-messenger astrophysical
constraints \citep{2009PhRvD..80j3510P} seems to indicate that
$\gamma$-ray photons from annihilation  of WIMP-like Cold Dark
Matter particles provide, at most, a minor contribution to the EGB.
Finally, the EGB could be mainly  consisting of a population of
numerous but faint $\gamma$-ray sources: normal, star-forming
galaxies are typical candidates that could make a substantial
contribution to the  EGB below 10 GeV
\citep{2002ApJ...575L...5P,2010arXiv1003.3647F}.

To tackle the problem of the  EGB nature one  can use different and
rather complementary methods (e.g. \citet{2009arXiv0912.1854H,
2009PhRvD..80h3504D}) that may serve as independent constraints:
\begin{itemize}

\item

Resolve the largest possible fraction of the EGB into individual
sources. If the EGB were mainly contributed by a population of rare,
bright objects, then Fermi-LAT will eventually be able to resolve a
significant fraction of this radiation and to disentangle possible
multiple components. If the EGB were mainly contributed by a
population of common, faint objects, then only a few of them will be
resolved by Fermi-LAT (see e.g. \citet{2001ApJ...558...63P} for the
case of star-forming galaxies). Finally, if the EGB were
significantly contributed by DM annihilations, then,  the Fermi-LAT
might be able to resolve only some individual sources associated to
foreground dark Galactic subhalos, depending on the precise nature
of the DM particle \citep{2008MNRAS.384.1627P,2009arXiv0908.0195P}.

\item

Compare the observed EGB energy spectrum with model predictions
based on the luminosity function of some classes of objects and
their energy spectra (e.g. \citet{2010arXiv1003.3647F}).

\item

Analyze the one- or two-point statistics of the observed EGB photon
counts to distinguish the contribution of clumpy components,
typically associated to individual sources, to that of a diffuse
component
\citep{2009JCAP...07..007L,2010arXiv1003.3647F,2010arXiv1006.2399B,2010MNRAS.405.1777S}.

\item

Analyze the angular correlation properties of the EGB and compare it
with those of a population of $\gamma$-ray emitting objects like
blazars \citep{2007PhRvD..75f3519A,2007MNRAS.376.1635A}, galaxy
clusters \citep{2007ApJ...667L...1M,2007PhRvD..75f3519A} type Ia
supernovae \citep{2004ApJ...614...37Z} star-forming galaxies
\citep{2009MNRAS.400.2122A} and DM halos
\citep{2006PhRvD..73b3521A,2007PhRvD..75f3519A,2007JCAP...04..013C,
2007JCAP...06..013H,2008PhRvD..77l3518C,2008JCAP...10..040S,
2009PhRvD..79d3521T,2009PhRvD..80b3518F,2009PhRvD..80b3520A,2010arXiv1005.0843C}.

\end{itemize}

In this work we follow the last approach and estimate the two-point
angular auto-correlation function (ACF) of the Fermi-LAT EGB  to
identify those characteristic features that can be related to the
presence of a well defined population of objects. Due to its
isotropy, the ACF signal of the EGB is expected to be weak. The
signal can be amplified by cross-correlating the EGB with the
angular distribution of the supposed EGB sources, since the number
of contributing sources  is typically large. For this reason we
cross-correlate the EGB with different catalogs of extragalactic
objects that can contribute to the EGB.

Our approach is very similar to that of \citet{2009MNRAS.400.2122A}
but our analysis is more comprehensive in two aspects. First of all,
since our aim is to identify the contributors to the EGB among a
variety of possible candidates we cross-correlate the EGB maps with
several different objects' catalogs (SDSS-DR6 quasars
\citep{2009ApJS..180...67R}, SDSS-DR6  Luminous Red Galaxies
\citep{2008arXiv0812.3831A}, NVSS radio galaxies
\citep{2002MNRAS.337..993B}) in addition to the 2MASS star-forming
galaxies \citep{2000AJ....120..298J} considered by
\citet{2009MNRAS.400.2122A}. Second of all, not only we do provide
theoretical predictions of the expected ACF and two-point
cross-correlation function (CCF), but we actually estimate them from
the 21-month Fermi-LAT data. Finally, we perform our analysis in
configuration space (although we also show the angular power spectra
for completeness) and consider energies $ E>1 $ GeV,  slightly
higher than in \citet{2009MNRAS.400.2122A}, to reduce the
contamination of the signal due to the Galactic foreground.

{In our work we also pursue another important goal: we attempt to
detect the Integrated Sacks-Wolfe (ISW) signal
\citep{1967ApJ...147...73S}. This effect  is related to the
variation over time of the gravitational potential $\Phi$ which
arises at late cosmological times when when dark energy or curvature
become important, and induces additional anisotropies in the Cosmic
Microwave Background (CMB) at large angular scales. The effect is
instead absent during the matter dominated era when the
gravitational potential is constant, and, for this reason, is a
potentially powerful probe for dark energy. Successful searches for
the ISW effect have been performed in the past by cross-correlating
the CMB maps with the large scale structures probes that we have
already mentioned (NVSS radio galaxies
\citep{2004ApJ...608...10N,2008MNRAS.386.2161R,2010JCAP...08..013X},
2MASS galaxies \citep{2004PhRvD..69h3524A,2007MNRAS.377.1085R}, SDSS
quasars \citep{2006PhRvD..74f3520G,2009JCAP...09..003X}), SDSS
galaxies \citep{2006MNRAS.372L..23C,2007MNRAS.381.1347C} and with
combinations of different tracers
\citep{2008PhRvD..77l3520G,2008PhRvD..78d3519H,2008PhRvD..78d3520H}.
Besides the above surveys, also the EGB sources are a probe of the
large scale structures, and cover a large fraction of the sky. Thus,
we do expect some ISW signal which could be detected through the
cross-correlation between Fermi and CMB data. Therefore, in this
paper we use the cross-correlation between the 21-month Fermi-LAT
EGB maps with the WMAP7 maps \citep{2010arXiv1001.4538K} of the CMB
to look for the ISW signal. }

Our analysis is also similar to that of \citet{1998NewA....3..275B}
and \citet{2004Natur.427...45B} since, like in their case, we also
cross-correlate two diffuse signals. In our case it is CMB versus
EGB whereas in their case the CMB is cross-correlated with hard
X-ray background.  However, for us, due to the large errors caused
by a smaller number of photons, we do not expect to be able to put
useful constraints on the cosmological constant or on the bias of
$\gamma$-ray sources.  Instead, we will use the cross-correlation
analysis to constrain the nature of the EGB in the framework of the
concordance Cold Dark Matter model with a cosmological constant
($\Lambda$CDM). Here we adopt the best-fit WMAP model
\citep{2010arXiv1001.4538K} in which $\Omega_{\rm b} h^2 = 0.02267$,
$\Omega_{\rm c} h^2 = 0.1131$, $\tau= 0.084$, $h = 0.705$, $A_{\rm
s} = 2.15\times 10^{-9}$ at $k_0=0.05$ Mpc$^{-1}$, and $n_{\rm s} =
0.968$.

The layout of the paper is as follows: in Section~\ref{sec:theory}
we briefly review the theoretical description of the
cross-correlation analysis, including the ISW signal. In
Section~\ref{sec:maps} we present the various maps used to compute
the two-point ACF and CCF. We describe the statistical estimator
used to evaluate the auto- and cross-correlation functions in
Section~\ref{sec:corranalysis}. The results of the cross-correlation
analysis are presented in Section~\ref{sec:results}, analyzed in
Section~\ref{sec:analysis} and discussed in
Section~\ref{sec:discussions}. The conclusions are presented in
Section~\ref{sec:conclusions}.

%%%%%%%%%%%%%%%%%%%%%%%%%%%%%%%%%%%%%%%%%%%%%%%%%%%%%%%%%%%%%%%%%%%%%%%%%%%%%%%%%%%%%%%%%%%%%%%%%%%%%%%%%%%%

\section{Theory}
\label{sec:theory}

Our theoretical formulation of the  mean $\gamma$-ray emission
contributed by unresolved sources, their auto-correlation and
cross-correlation with the angular distribution of different types
of extragalactic objects follows that of
\citet{2009MNRAS.400.2122A}. The treatment of the cross-correlation
with the CMB maps and the related ISW theory follows that of
\citet{1998NewA....3..275B} and \citet{2009JCAP...09..003X}.

\subsection{Mean Intensity}
\label{sec:meansignal}

The mean differential $\gamma$-ray \emph{energy flux} due to a
population of sources $j$ characterized by a $\gamma$-ray luminosity
function $\Phi_j(L_{\gamma},z)$ is:

\begin{equation}
\frac{dI_j}{dE}=\frac{c}{4 \pi} \int \left[ \int_{L_{\rm
MIN}}^{L_{\rm MAX}(z)} \Phi_j(L_{\gamma},(1+z)E,z)L_{\gamma}
dL_{\gamma} \right] \frac{dz}{(1+z)H(z)}~, \label{eq:dide}
\end{equation}
where $E$ is the energy, $z$ is the redshift,
$H(z)=H_0[(1+z)^3\Omega_M+\Omega_{\Lambda}]$ is the expansion
history of a flat universe with a cosmological constant $\Lambda$
and the $(1+z)^{-1}$ term represents cosmological dimming of the
photon energy. {$L_{\gamma}$  is the source luminosity, given
throughout the paper  in erg~s$^{-1}$, while we measure  $dI_j/dE$
in erg~cm$^{-2}$s$^{-1}$sr$^{-1}$. }  $\rho_{\gamma}(z) \equiv \int
\Phi(L)L dL$ represents the comoving $\gamma$-ray luminosity density
of the sources at redshift $z$. The integration limits are set by
requiring that the EGB consists entirely of sources below the
\emph{photon} flux detection limit  $S_{\rm lim}$
(ph~cm$^{-2}$s$^{-1}$) In this case $L_{\rm MAX}(z)=4 \pi d_L^2(z)
F_{\rm lim} (1+z)^{-2+\Gamma_j}$ \citep{2009MNRAS.396L.105G}, where
$d_L$ is the luminosity distance in the adopted cosmology and
$\Gamma_j$ is the \emph{photon} index of the source population
energy spectrum, assumed to be a power law (see also below). $F_{\rm
lim}$ is the \emph{energy} flux detection limit
(erg~cm$^{-2}$s$^{-1}$) , which is related to the photon flux
$S_{\rm lim}$ through $F_{\rm lim}=S_{\rm lim} E_{t}\times
(1-\Gamma_j)/(2-\Gamma_j)$, where $E_{t}$ is the energy threshold of
integration, typically 100 MeV. $L_{\rm MIN}$ can be, in principle,
set to zero if the total luminosity density  $\rho_{\gamma}(z)$ were
convergent. Unluckily, for the source classes discussed below the
extrapolation of the luminosity function to small values of $L_{\rm
MIN}$ gives a divergent total luminosity.  However, an effective
$L_{\rm MIN}$, if not given {\it a priori}, can be nonetheless
computed by extrapolating the observed $\Phi_{\rm j}(L)$ and by
requiring that the population of sources $j$ contribute a fraction
$f_{\rm j}$ of the total EGB:
\begin{equation}
f_{\rm j} \int_{100}^{+\infty} \frac{dI_{\rm EGB}}{dE}
\frac{dE}{E}\equiv\int_{100}^{+\infty} \frac{dI_{\rm j}}{dE}
\frac{dE}{E}\ =\int_0^{S_{\rm lim}} \frac{dN_{\rm j}}{dS_{100}}
S_{100} dS_{100}, \label{eq:fegb}
\end{equation}
where $dN_{\rm j}/dS_{100}$ represents the differential counts of
$j$ sources {and the factor of $E^{-1}$ accounts for the units of
the $dN_{\rm j}/dS_{100}$ function  typically given in terms of
photon flux  rather than energy flux\citep{2010ApJ...720..435A}. }
The conventional threshold of 100 MeV has been chosen to define the
integral flux above 100 MeV $S_{100}$. This expression assumes that
the detection efficiency can be modeled as a step function. In fact
\cite{2010ApJ...720..435A} found that the Fermi-LAT detection
efficiency for $E>100$ MeV drops as $S^{-2}$ below $S_{\rm
lim}=3\times 10^{-8} \ {\rm ph} \ {\rm cm}^{-2} {\rm s}^{-1}$. We
verified that, varying $S_{\rm lim}$ in the reasonable interval
$10^{-8}$-$10^{-7} \ {\rm ph} \ {\rm cm}^{-2} {\rm s}^{-1}$, this
stepwise approximation does not affect significantly the resulting
redshift distribution of unresolved sources. From Eqs.~\ref{eq:dide}
and~\ref{eq:fegb} then it is possible to compute the mean
differential flux and the fraction $f_{\rm j}$ of the EGB emission
contributed by any class of unresolved sources once their luminosity
function and number counts are modeled theoretically or extrapolated
from observations.

In this work we will assume that unresolved sources have power law
energy spectra $I(E)dE \propto E^{1-\Gamma_j}dE$ and photon index
$\Gamma_j>1$. In this case the energy dependence drops out of the
integral in Eq.~\ref{eq:dide} and the integrated energy flux
becomes:
\begin{eqnarray}
\nonumber
I_j(>E)= \int^{\infty}_{E} \frac{dI_j}{dE} dE = \ \ \ \ \ \ \ \ \ \ \ \ \ \ \ \ \ \ \ \ \ \ \ \ \ \ \ \ \ \ \ \ \ \ \ \ \ \ \
\ \ \ \ \ \ \ \ \ \  &   \\
= \frac{cE^{2-\Gamma_j}}{4\pi}  \int  \left[ \int_{L_{\rm MIN}}^{L_{\rm MAX}(z)} \Phi_j(L_{\gamma},z)L_{\gamma} dL_{\gamma} \right]
\frac{(1+z)^{-\Gamma_j}}{H(z)}dz  \;. & \
\label{eq:integratedflux}
\end{eqnarray}

\begin{figure}
\centering
\epsfig{file=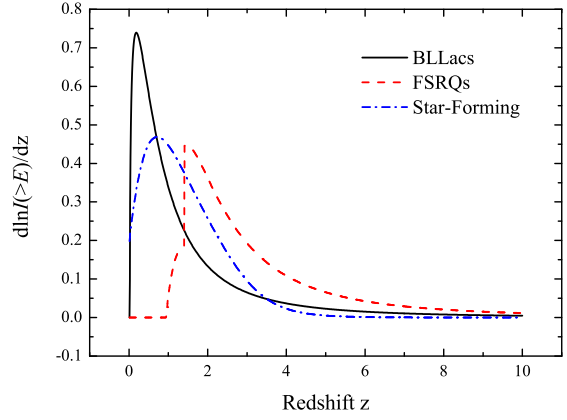, angle=0, width=0.52 \textwidth}
\caption{Normalized $\gamma$-ray flux per unit redshift $d\ln
I(>E)/dz$ as a function of $z$ for  three different source classes:
FSRQs (red, dashed), BLLacs (black, continuous) and star-forming
galaxies (blue, dot-dashed).} \label{fig:zdist1}
\end{figure}

In this paper we deal with maps of photon counts rather than energy
flux; the photon flux (above energy $E$) being simply
\mbox{$(2-\Gamma_j)/(1-\Gamma_j)\times I_j(>E)/E$}. We will
considered integrated fluxes with  three energy thresholds:
$I(>E={\rm 1 GeV})$,  $I(>E={\rm 3 GeV})$ and $I(>E={\rm 30 GeV})$
and three possible contributors to the EGB: two types of blazars,
Flat Spectrum Radio Quasars (FSRQs) and BL Lacertae (BLLacs), and
star-forming galaxies. The main characteristics of these three
populations, summarized in Table~\ref{tab:tab1}, are:

\begin{enumerate}
\item {\it FSRQs} represent a sub-class of blazars, i.e. AGNs with a
relativistic jet pointing close to the line of sight. For the
classification of blazars as FSRQs we rely on the criteria adopted
by \citet{2009ApJ...700..597A}. These authors also compute the FSRQs
luminosity function for $E>100$ MeV in three different redshift
intervals, and we adopt their description under the assumption that
its shape does not depend on the energy band. The FSRQs number
counts have been measured by \citet{2010ApJ...720..435A} from a
larger sample of objects. The slope of the counts flattens in the
faint end, implying that \mbox{FSRQs} at most contribute to $f_{\rm
j}\sim 25$ \% of the EGB. Their spectra are steeper than those of
BLLacs, with an average photon index of $\Gamma\sim2.47$. To model
the FSRQs contribution to EGB we have derived an  $f_{\rm j}$
estimate from the log$N$-log$S$ of \citet{2010ApJ...720..435A}. We
then use the luminosity function and, as anticipated, we enforce an
{\it effective} luminosity cut  $L_{\rm MIN}$. The calculation gives
$L_{\rm MIN} = 0.2 \times 10^{48}$erg s$^{-1}$.  The resulting
redshift distribution of the flux is shown in Fig. \ref{fig:zdist1},
with the apparent piecewise behavior resulting from the different
luminosity functions in the different redshift bins. We stress that
the value of $L_{\rm MIN}$ derived in this way is not fully
consistent with the minimum FSRQ luminosity measured by
\citet{2009ApJ...700..597A}. This affects the redshift distribution
in the lowest redshift bin, which however makes only a subdominant
contribution to the total flux. Given the large uncertainties in the
modeling of the FSRQs luminosity functions and their counts,  this
is likely a reasonable level of approximation. Finally, for these
sources we adopt the redshift-dependent AGN biasing function
proposed by \citet{2009MNRAS.396..423B} in the framework of the
semi-analytic models of AGN-black holes co-evolution:
$b_{\gamma}(z)=0.42+0.04(1+z)+0.25(1+z)^2$.

\item {\it BLLacs} are another sub-class of blazars, on average less-bright
than FSRQs. As in the previous case we adopt the BLLacs luminosity
function measured by \citet{2009ApJ...700..597A}, the number counts
determined by \citet{2010ApJ...720..435A} and enforce an effective
value of $L_{\rm MIN}$ to reconcile the two predictions. The
corresponding contribution to the EGB is $f_{\rm j}\sim 12$ \%,
while $L_{\rm MIN} = 6 \times 10^{43}$erg s$^{-1}$. Finally, we use
the average photon index measured by \citet{2010ApJ...720..435A},
i.e. $\Gamma \sim 2.2$ and assume that BLLacs  and FSRQs have the
same bias factor $b(z)$. In this case the luminosity function is
compatible with no-evolution in the different redshift bins and the
corresponding redshift distribution in Fig. \ref{fig:zdist1} has no
piecewise behavior.

\item {\it Star-forming galaxies} are fainter and much more common than
blazars. In fact, it has been proposed  that they alone could
account for the EGB fraction which is not contributed by unresolved
blazars. \citet{2010arXiv1003.3647F} have shown that under the
assumption of an Euclidean faint-end slope for the source counts,
their contribution to the EGB can be as large as $f_{\rm j}=70$ \%,
an assumption that we will also adopt in our analysis. Since they
are very faint and  difficult to resolve, their luminosity function
cannot be determined experimentally but needs to be  modeled
theoretically. \citet{2009MNRAS.400.2122A} have proposed a model in
which the luminosity of each single source scales with the
star-formation rate and the gas mass fraction. Since in this model
the $\gamma$-ray emissivity is rescaled from that of the Milky Way,
the underlying assumption is that most of the $\gamma$-ray photons
are emitted from de-evolved versions of our own Galaxy.  Current
theoretical uncertainties and weak observational constraints do not
allow to discriminate among the simple model proposed by
\citet{2009MNRAS.400.2122A} and the more recent (and more
sophisticated) one presented by \citet{2010arXiv1003.3647F}. For
this reason in this paper we have decided to adopt the first one.
The energy spectrum of star-forming galaxies is characterized by a
strong pionic peak at $E\sim 0.2$ GeV, a feature that also
determines the energy dependence of their  contribution to the EGB.
In the energy range we are interested in ($E>1$ GeV), their energy
spectrum is  fairly well approximated by a power law behavior with
$\Gamma \sim 2.475$ \citep{2010arXiv1003.3647F} that allows us to
use Eq.~\ref{eq:integratedflux}. Finally, we assume that
star-forming galaxies are unbiased, i.e. $b_S=1$, as suggested by
observations \citep{2004PhRvD..69h3524A}. The results of the auto-
and cross-correlation studies do not change significantly as long as
the bias of these galaxies is close to unity
\citep{2009MNRAS.400.2122A}.

\end{enumerate}

In Fig.~\ref{fig:zdist1} we show the redshift dependence of the
normalized energy flux per unit redshift $d\ln I(>E)/dz$ for the
three proposed EGB sources: FSRQs (red, dashed), BLLacs (black,
continuous) and star-forming galaxies (blue, dot-dashed). This
function, proportional to the integrand in
Eq.~\ref{eq:integratedflux}, represents the contribution to the EGB
from the sources in a specific redshift range. In the BLLacs
scenario, the contribution to the EGB signal is relatively local,
i.e. produced by a population of faint, nearby sources. {In the
FSRQs case the signal mostly comes from $z\ge 2$ and drops to zero
at $z<0.9$. This reflects the fact that FSRQs are rare bright
objects which, for $z<0.9$, would have $L_{max}<L_{MIN}$ in
Eq.\ref{eq:dide}, i.e. would be above detection threshold and
removed from the map. As a consequence, the cross correlation of the
FSRQs signal with catalogs of objects whose number density peaks at
low redhifts is expected to be zero, as we shall se in Section
\ref{sec:results}. } Star-forming galaxies represent an intermediate
case in which the signal is produced over a relatively broad
redshift range around $z\sim1$.

\begin{table}
%[ht]
\begin{center}
\caption{ Investigated EGB contributors. All values refer to $E>3$
GeV and $S<S_{lim}$. The luminosity function, $\frac{dN}{dS}$ and
spectral parameters for FSRQs and BLLacs are taken from
\citet{2009ApJ...700..597A} and \citet{2010ApJ...720..435A}
\label{tab:tab1}.}
\begin{tabular}{|c|c|l|c|c|}
\hline\hline
Source Type &
$f_{\rm j}$ &
$\Phi(L_{\gamma},z) \propto L^{-\alpha}$ &
$\frac{dN}{dS} \propto S^{-\beta}$  &
$I(E)\propto E^{-\Gamma}$  \\
\hline
& & $\alpha=1.57 \ \  z<0.9 $ & &   \\
FSRQs  & 25 \% &
$\alpha=2.45 \ \ z=[0.9,1.4]$ &
$\beta=1.72$ &
$\Gamma=2.47$ \\
&& $\alpha= 2.58 \ \  z>1.4 $ && \\
BLLacs  & 12 \% & $\alpha=2.23  \ \  z \ge 0$ &$\beta=0.70$ & $\Gamma=2.20$ \\
Star-form. & 70 \% & See Text & $\beta=2.5$ & $\Gamma=2.45$ \\
\hline
\end{tabular}
\end{center}
\end{table}

\subsection{Fluctuations in the $\gamma$-ray flux}
\label{sec:fluctuations}

To compute the predicted auto- and cross-correlation signals we need
to model the  fluctuations of the $\gamma$-ray flux. These
fluctuations arise from local deviations from  the $\gamma$-ray
luminosity density  $\rho_{\gamma}(z)$ that we assume to be
proportional to the deviations from the mean number density of
sources $n_{\gamma}(z)\equiv\int  \Phi(L) dL$:
 \begin{equation}
\delta_{\gamma}(z,{\bf x}) \equiv \frac{\rho_{\gamma}(z,{\bf
x})-\rho_{\gamma}(z)} {\rho_{\gamma}(z)}= \frac{n_{\gamma}(z,{\bf
x})-n_{\gamma}(z)}{n_{\gamma}(z)} \equiv \delta_{n_{\gamma}}(z, {\bf
x})~.
\end{equation}
We also assume that the $\gamma$-ray sources trace the underlying
fluctuations in the mass density according to some linear biasing
prescription that may depend on the redshift:
\begin{equation}
\delta_{n_{\gamma}}(z,{\bf x})\equiv b_{\gamma}(z)\delta_m(z,{\bf
x})= b_{\gamma}(z) \frac{\rho_m(z,{\bf x})-\rho_m(z)} {\rho_m(z)}~,
\end{equation}
where $\rho_m$ indicates the mass density and $b_{\gamma}(z)$ is
called the biasing function.

Putting all together, the expected fluctuation in $\gamma$-ray
energy flux is:
%\frac{\int (1+z)^{1-\Gamma}H(z)^{-1} \rho_{\$\gamma$}[z]\delta_{\$\gamma$}[z,{\bf x}] dz}
%        {\int (1+z)^{1-\Gamma}H(z)^{-1} \rho_{\$\gamma$}[z]dz}=
%\end{equation}
%\begin{equation}

\begin{equation}
\delta I( {\bf n} ) \equiv \frac{I({\bf n}) -   I}{I}= \frac{\int
(1+z)^{-\Gamma}H(z)^{-1} \rho_{\gamma}(z)b_{\gamma}(z)
\delta_m(z,{\bf x})dz} {\int (1+z)^{-\Gamma}H(z)^{-1}
\rho_{\gamma}(z) dz}~,
\end{equation}
where $I \equiv I(>E)$ indicates the $\gamma$-ray mean flux and $I(
{\bf n} )\equiv I( >E,{\bf n} )$ is the energy flux along the
generic direction ${\bf n}$.

\subsection{Two-point Angular Correlation Function and Spectrum}
\label{sec:correlation}

We can now compute the expressions for the two-point ACF of the EGB
fluctuation field $E$ and the  two-point CCF between $E$ and another
fluctuation field obtained from a source catalog, $S$.

The general expression for the two-point angular correlation is
\begin{equation}
\left \langle  \delta I( {\bf n_1} ) \delta  J( {\bf n_2} ) \right
\rangle = \sum_l\frac{2l+1}{4\pi}C_l^{I,J} P_l[cos(\theta)]~,
\label{eq:2point}
\end{equation}
where $I$ and $J$ are the two fields and the angular spectrum is
given by:
\begin{equation}
C_l^{I,J} =\frac{2}{\pi} \int k^2 P(k) [G_l^I(k)] [G_J^I(k)]  dk~.
\label{eq:angularspectrum}
\end{equation}
and $P(k)$ is the present-day power spectrum.

For example, for the EGB ACF one has $I=J=E$ with
\begin{equation}
G_l^E(k)=\frac{\int (1+z)^{-\Gamma}H(z)^{-1}
\rho_{\gamma}(z)b_{\gamma}(z) D(z) j_l[k \chi(z)]dz} {\int
(1+z)^{-\Gamma}H(z)^{-1} \rho_{\gamma}(z) dz}~, \label{eq:autocorr}
\end{equation}
where $j_l[k \chi(z)]$ are spherical Bessel functions, $D(z)$ is the
linear growth factor of density fluctuations and $\chi(z)$ is the
comoving distance to redshift $z$. In our analysis, we use the
public code {\tt CAMB} \citep{2002PhRvD..66j3511L} to generate the
linear power spectrum of density fluctuations and the {\tt Halofit}
\citep{2003MNRAS.341.1311S} built-in routine for non-linear
correction to obtain the fully-evolved, nonlinear matter power
spectrum $P(k)$ at any epoch. We note that non-linear corrections do
not affect our results significantly.

In case of cross-correlation with a fluctuation field of discrete,
unresolved sources one has $I=E$, $J=S$  and
\begin{equation}
G_l^S(k)=\int \frac{dN(z)}{dz} b_S(z) D(z) j_l[k \chi(z)]dz~,
\label{eq:crossocorr}
\end{equation}
where $dN(z)/dz$ and $ b_{\gamma}(z)$ represents the redshift
distribution and the bias factor of the sources that do not
necessarily coincide with the $\gamma$-ray sources.

Finally, if one cross correlates EGB with another diffuse signal,
like the temperature fluctuation field obtained from the CMB maps
($I=E$, $J=T$) then
\begin{equation}
G_l^{T}(k)=-2\int \frac{d\Phi(k)}{dz}j_l[k(z)]dz~, \label{eq:isw}
\end{equation}
where $\Phi$ represents the gravitational potential.

In this case the cross-correlation signal in Eq. \ref{eq:2point}
represents the ISW effect, expected if the EGB were contributed by
sources that trace the underlying mass distribution.

In Section \ref{sec:results} we will use Eqs.~\ref{eq:2point},
\ref{eq:angularspectrum}, \ref{eq:autocorr}, \ref{eq:crossocorr} and
\ref{eq:isw} to predict the expected auto- and cross-correlation
signal that will be compared with data.

\section{Maps}
\label{sec:maps}

In this section we describe the various maps (residual isotropic
Fermi-LAT maps, WMAP7 CMB, SDSS QSOs, SDSS, LRGs, 2MASS galaxies,
NVSS radio galaxies) that will be used for the auto- and
cross-correlation analysis.

\subsection{Fermi-LAT maps}
\label{sec:fermimaps}

The Large Area Telescope (LAT) is the primary instrument onboard the
Fermi Gamma-ray Space Telescope launched in June 2008
\citep{2009ApJ...697.1071A}. It is a pair-conversion telescope
covering the energy range between 20 MeV and 300 GeV. Due to its
excellent angular resolution ($\sim 0.1^{\circ}$ above 10 GeV), very
large field of view ($\sim 2.4$ sr) and efficient rejection of
background from charged particles, it represents a key experiment
for $\gamma$-ray astronomy. Fermi-LAT is continuously scanning the
sky in survey mode, providing a complete image of the sky every 3
hours.

The data gathered by the telescope are available online
\footnote{http://fermi.gsfc.nasa.gov/ssc/data/access/}. For the
present analysis we used, unless otherwise specified, 21 months of
data. We used only events classified as {\it class 4}. {\it Class 4}
events form the dataset with the lowest level of CR background
contamination currently available for LAT data analysis. Details on
this event classification are described in
\citet{2010PhRvL.104j1101A}. We also use events labeled as {\it
class 3} (cf. \citet{2009ApJ...697.1071A}) for consistency checks
\footnote{{\it Class 4} events are also referred to as ``{\it
dataclean}''  class while the union of {\it class 3} and {\it class
4} events is also referred to as  ``{\it diffuse}'' class. See
http://fermi.gsfc.nasa.gov/ssc/data/analysis/documentation/Cicerone
\!\!\!\!\! /Cicerone\_Data\_Exploration/Data\_preparation.html}.
{\it Class 3}   events have a larger residual contamination. We
apply a cut of $100^\circ$ on the zenith angle, $52^\circ$ on the
satellite rocking angle and $65^\circ$ on the inclination angle to
reduce the contamination  from Earth albedo.  The counts then have
been  pixelized in the HEALPix format \citep{2005ApJ...622..759G}
with $N_{\rm side}=64$, corresponding to $N_{\rm pix} = 49152$
pixels with an angular size $0.92^\circ\times0.92^\circ$. The
majority of these events come from the resolved point sources
\citep{2010ApJS..188..405A} and from the Galactic diffuse emission
due to $\gamma$-rays produced in the interaction of cosmic rays with
the interstellar medium. Both need to be removed to extract the EGB
signal.

To account for the Galactic diffuse foreground we  adopted  the
model \verb"gll_iem_v02.fit" \footnote{A more detailed description
can be found at
http://fermi.gsfc.nasa.gov/ssc/data/access/lat/BackgroundModels.html}.
In the following this will be referred to as {\it V1}. The model is
based on fits of the LAT data to templates of the HI and CO gas
distribution in the Galaxy as well as an Inverse Compton model
obtained with the GALPROP code
\footnote{http://galprop.stanford.edu/} and a further template for
the Loop-I region \citep{2009arXiv0912.3478C}. The model, based on 1
year of data taking by Fermi-LAT, describes well the Galactic
diffuse emission over the sky, with some exception, most prominently
in the regions which have been associated with giant gamma-ray lobes
\citep{2010arXiv1005.5480S}.  We will thus also use a preliminary
version of a refined model in development, which, further, includes
templates specifically for these structures, sensibly reducing the
residuals in these regions.   This is model {\it V2}. In the next
section we check the impact of these models  on the ACFs-CCFs and
compare the results with those obtained by masking out the areas
where the Bubble and Loop-I emission are more prominent. The mask is
obtained from the tabulated contours given in
\citep{2010arXiv1005.5480S} excluding the whole region enclosed by
Loop-I in the North Galactic sky. In addition, for both models, we
masked out a 20$^{\circ}$ strip above and below the Galactic plane
where the EGB signal is largely subdominant. And in order to check
the robustness of our results, we vary the strip mask width from
$b=20^{\circ}$ to $b=50^{\circ}$, in steps of $\Delta b=10^{\circ}$.

\begin{figure}
\centering \epsfig{file=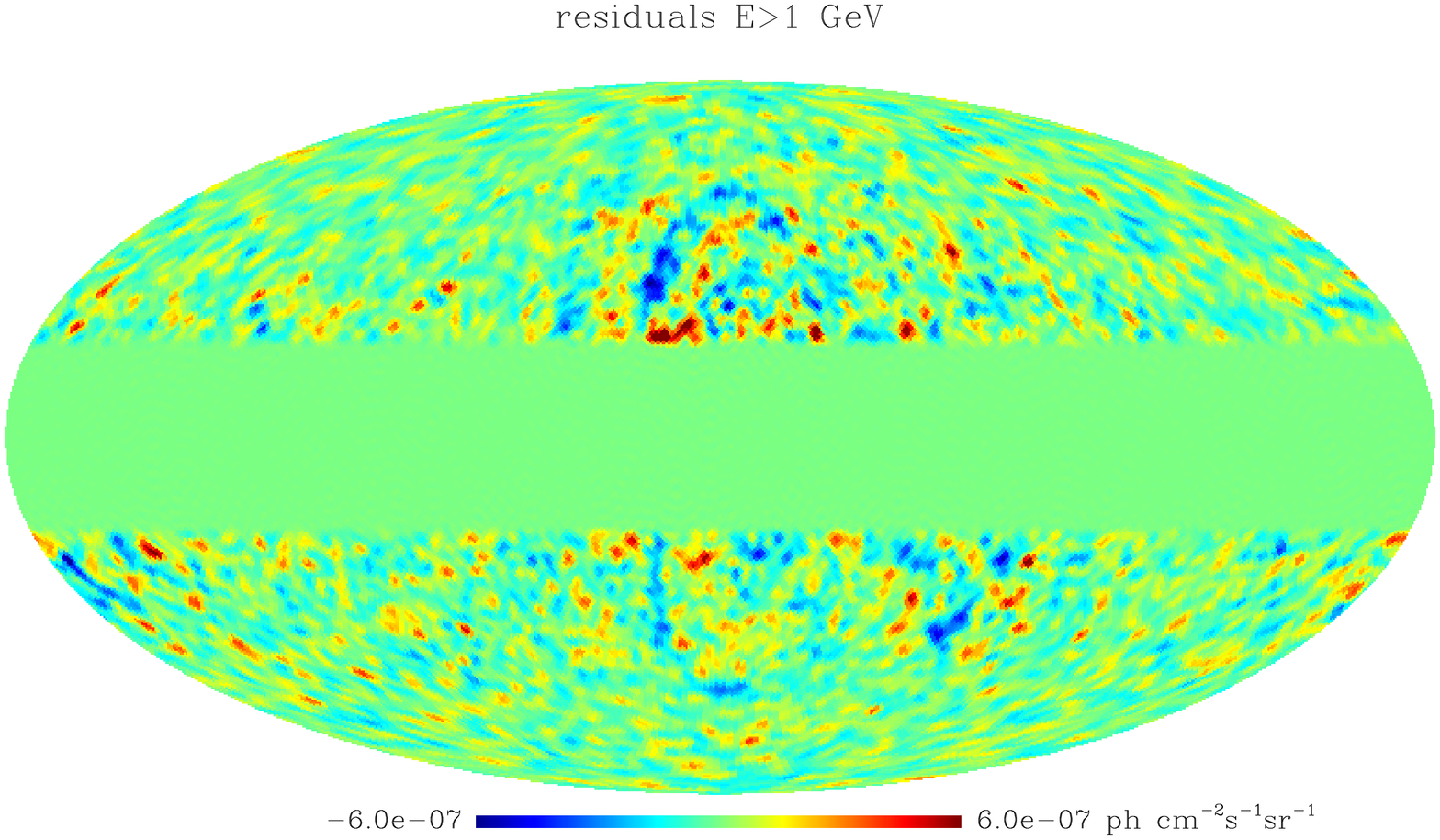, angle=90, width=0.45
\textwidth} \epsfig{file=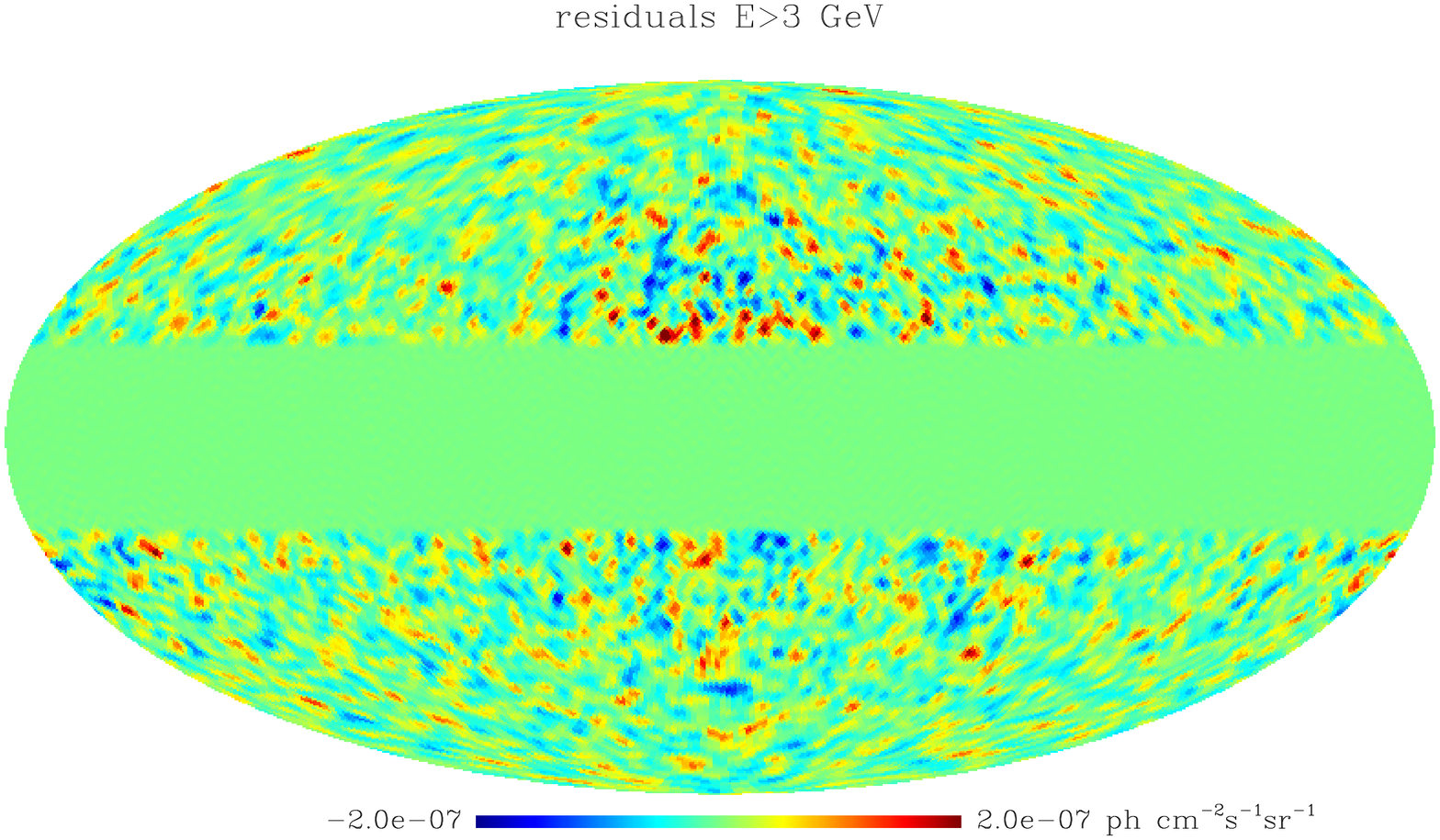, angle=90,
width=0.45 \textwidth} \epsfig{file=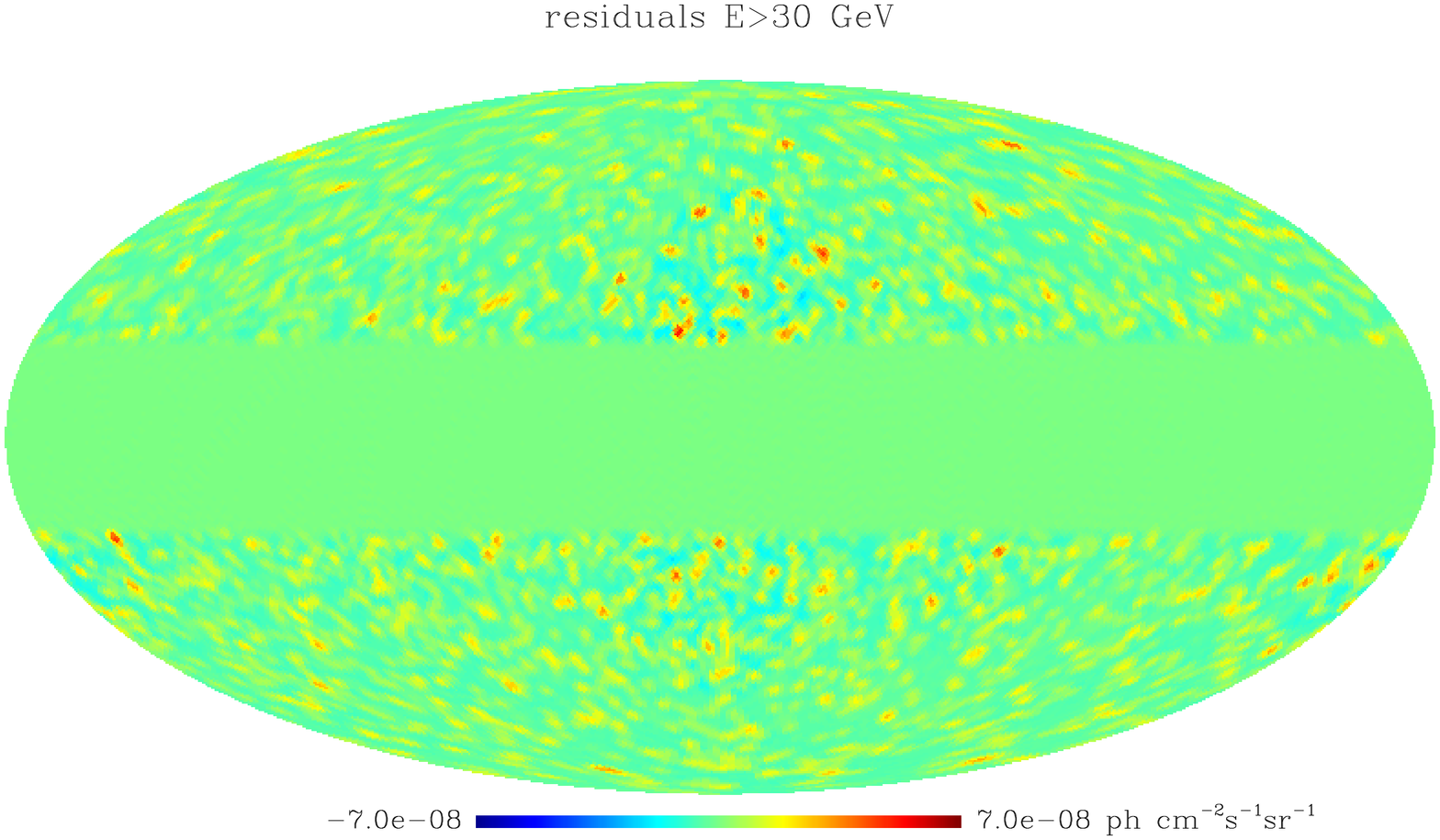,
angle=90, width=0.45 \textwidth} \epsfig{file=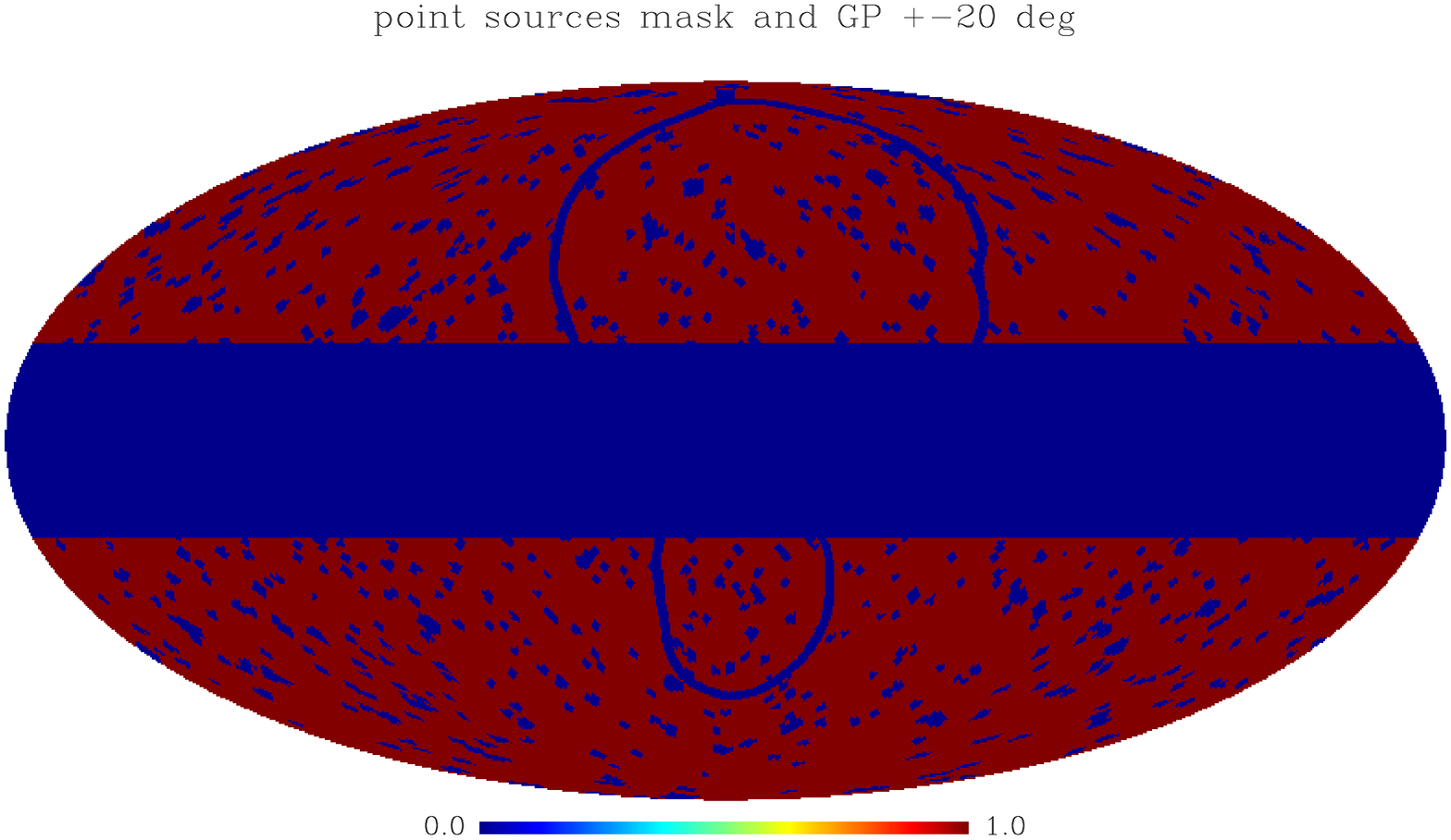,
angle=90, width=0.45 \textwidth} \caption{Residual gamma-ray
emission obtained from analyzing 21 months of Fermi-LAT data for
energies $E>1$ GeV (top), $E>3$ GeV (second from the top),  $E>30$
GeV (second from the bottom) and the mask covering the point sources
in the 1 year Fermi-LAT point sources catalog and the region with
$|b|<20^{\circ}$ (bottom panel). The two contours in the  mask
indicate the regions which are further removed when the Lobes/Loop-I
mask is applied.  The residual maps have been smoothed with a
$2^\circ$ Gaussian filter to remove small scale Poisson noise. All
maps are in Galactic coordinates and have a resolution $N_{\rm
side}=64$ ($\sim0.92^{\circ}$). All Fermi-LAT maps have been cleaned
by removing the $V2$ Galactic model and all multipoles with $\ell
\le 10$ (see text).} \label{fig:Fermi_maps}
\end{figure}

Point sources in the 1 year Fermi-LAT point-source catalogue 1FGL
\citep{2010ApJS..188..405A} are also masked as described in
\citet{2010arXiv1005.0843C}. The point source masking is adaptive,
in the sense that more intense sources, more likely to bias the
results, are masked with a larger circle (up to 2-3 degree radius),
while fainter ones are masked with a smaller circle down to about 1
degree (see Fig.~\ref{fig:Fermi_maps}).

To subtract the Galactic foreground from the data we have first
generated the energy dependent exposure  maps with the
\verb"gtexpcube" routine available in the public Fermi-LAT science
tools \footnote{See http://fermi.gsfc.nasa.gov/ssc/}. Version p6v3
{\it class 4} of the LAT instrument response functions (IRFs) and
the same cuts used for the event selections were used for the
exposure map generation. Maps of expected foreground counts from our
two adopted Galactic diffuse emission models have been calculated by
applying the exposure maps to the models and convolving them with
the Point Spread Function (PSF) of the LAT averaged over the field
of view and the relevant energy range.  The  \verb"GaRDiAn" package
\citep{2008AIPC.1085..763A}   was used for this task. We thus
include explicitly  the PSF convolution, although we stress that in
the energy range considered the PSF width is  always smaller or at
most comparable to our minimum angular bin (approximately
1$^\circ$). However, further  checks of the effect of the PSF on the
ACF are considered  in the next section. After the subtraction, the
residuals should be dominated by the EGB signal and by the residual
isotropic instrumental background. We have produced three maps
containing events with energy $E>1$ GeV, $E>3$ GeV and $E>30$ GeV,
respectively. The energy cut was performed to reduce the chance of
residual contamination from Galactic emission and to evaluate
different trade-offs between clean maps and reasonably large
statistics.

Since models $V1$ and $V2$ are not perfect, residuals can contain
some spurious signal on large scales that may affect the correlation
studies. The impact on the CCF is small since these spurious
residuals are not expected to correlate with the extragalactic
signal, but might be significant on the ACF. For all the maps, we
thus always remove the residual monopole and dipole contribution.
This is  performed using the specific  HEALPix routine,
\verb"remove_dipole". In addition, we have implemented a more
aggressive cleaning procedure in which, besides removing the
monopole and dipole, we have expanded the residual maps into
spherical harmonics using the HEALPix tool \verb"anafast" and have
removed all the multipoles  up to $\ell=10$ that contribute to the
large scale signal. We will refer to the residual maps with all
$\ell <10$ multipoles removed as $\ell10$ maps whereas $\ell1$
indicates the maps in which only dipole an monopole have been
removed. The validation of this procedure and  testing for possible,
undesired systematic effects is described in the next section.

The three $\ell10$ residual counts maps are shown in Fig.
\ref{fig:Fermi_maps} together with our fiducial mask removing the
region within 20 degrees from the Galactic Plane and the point
sources. To better illustrate the fluctuation properties of these
maps, the mean has been removed, thus showing only fluctuations
around zero. Furthermore, the  maps have been smoothed with a
$2^\circ$ Gaussian filter to remove small scale Poisson noise.

\subsection{Validation and checks}
\label{sec:validation}

\begin{figure}
%\centering
\hspace{-0.8cm}
\vspace{-0.3cm}
\epsfig{file=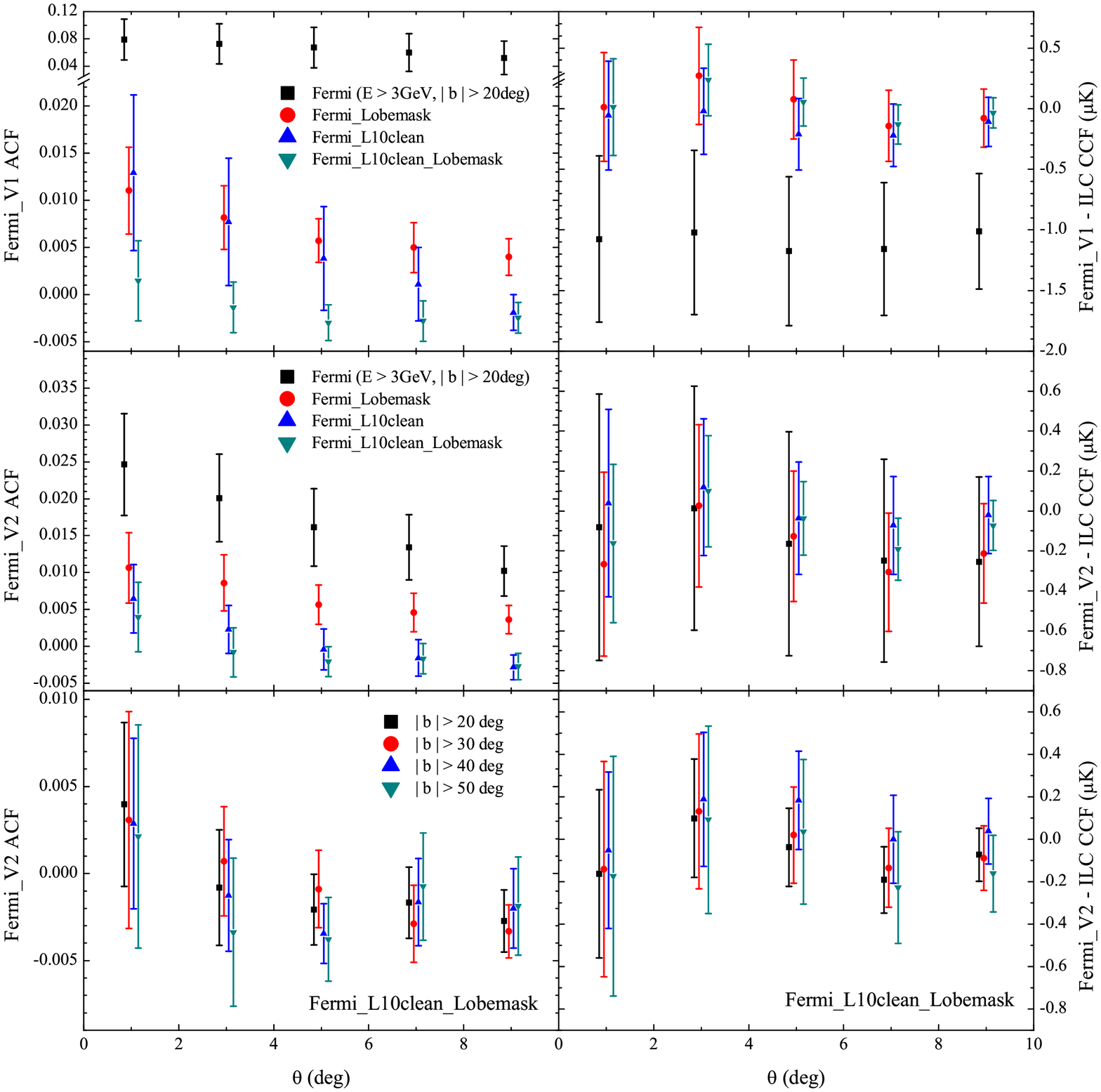, angle=0, width=0.55 \textwidth}
\vspace{-1.2cm}
%\epsfig{file=data_3Gev_acf_l10clean_lobes, angle=0, width=0.45 \textwidth}
%\vspace{-1cm}
%\epsfig{file=data_3Gev_ccf_20, angle=0, width=0.45 \textwidth}
%\vspace{-0.5cm}
%\epsfig{file=data_3Gev_ccf_l10clean_lobes, angle=0, width=0.45 \textwidth}
\caption{ACFs (left panels) and CCFs with the CMB (right panels) for
different cleaning procedures ($\ell10$ and $\ell1$, with/without
Bubbles/Loop-I mask) for the {\it V1} Galactic model (top panels),
{\it V2} Galactic model (middle panels), and for different Galactic
latitude cuts (bottom panels). All plots refer to Fermi  $E>3$ GeV
residual maps.} \label{fig:ACFCCFcheckse3GeV}
\end{figure}

The scope of this section is to describe the various tests that we
have performed to check and validate our cleaning procedure, to
reveal the presence of possible systematic effects,  to correct them
and estimate their impact on the ACF and CCF analysis. For brevity
we only show a subset of the results obtained, but we stress that
all robustness tests described in this section have been repeated
for each of the ACF/CCFs analyses presented in this paper.

\subsubsection{Validating the $\ell10$ cleaning procedure}
\label{sec:montecarlo}

To validate the $\ell10$ cleaning procedure we have applied this
cleaning technique to a set of mock data mimicking the
characteristics of the real 21-month Fermi-LAT data. The mock, Monte
Carlo-generated data-sets are simulated with the \verb"gtobssim"
routine using the \verb"p6_v3_diffuse" Instrument Response Functions
and including the  contribution from three signals {\it (i)} a
Galactic diffuse component generated using the $V1$ Galactic model,
{\it (ii)} an extragalactic diffuse and isotropic component (hence
with no intrinsic ACF) generated with the
\verb"isotropic_iem_v02.txt"
\footnote{http://fermi.gsfc.nasa.gov/ssc/data/access/lat/BackgroundModels.html}
model and {\it (iii)} a signal contributed by a population of AGNs
generated by sampling their observed log$N$-log$S$ distribution
\citep{2010ApJ...720..435A} of which we mask the 1200 brightest
ones. {No intrinsic clustering is assumed for the simulated AGNs
which are distributed across the sky randomly. } The same pipeline
described in the previous section is then applied to the mock
datasets to obtain the residual maps from which ACFs and CCFs with
the CMB are calculated. We observe that the resulting ACFs and CCFs
are generally nicely consistent with zero. Some systematic features
at the level of $1\sigma$ however appear as, mainly, a negative
offset of the ACF for the $\ell10$ maps. Indeed, we also observe
this offset for the ACF of real data. We thus  plot a gray region
around the measured ACF to indicate the presence of this systematic.

\subsubsection{Robustness to the cleaning procedure, Galactic model and
masking}

To test the robustness of our results with respect to the cleaning
procedure ($\ell1$ vs. $\ell10$), Galactic model ($V1$ vs. $V2$) and
Galactic mask ($|b|<20^{\circ}, \, 30^{\circ}, \, 40^{\circ}, \,
50^{\circ}$ and Bubbles/Loop-I cut) we have computed the residuals
of {real Fermi maps } using the different combinations of these
procedures at all energy cuts. Then we have computed and compared
the various ACFs and CCFs with all catalogs described in the
following section. An example of these tests is shown in
Fig.~\ref{fig:ACFCCFcheckse3GeV}. The different panels show the ACF
(left) and the CCF with the CMB (right) relative to $V2$ model and
$E>3$ GeV cut. {Top panels show the impact of the different cleaning
procedures for the $V1$ galactic model. Middle panels are relative
to the $V2$ galactic model. } Bottom panels show the impact of
different $|b|$ cuts.

The results of these tests can be summarized as the following: {\it
(i)} for the ACF analysis the Galactic model $V2$ performs
significantly better than $V1$ and therefore will be used throughout
this work. {\it (ii)} The ACF is quite sensitive to different
cleaning procedures and both the Bubbles/Loop-I mask and the
$\ell10$ cleaning are required to get  consistent results.  After
this cleaning, the auto-correlation signal is stable for Galactic
cuts $|b| \ge 20^{\circ}$. For these reasons, we will consider (and
show) only the ACFs computed using the $V2$ model with $\ell10$
cleaning, Bubbles/Loop-I mask and a Galactic cut $|b| = 20^{\circ}$.
{\it (iii)} The CCFs are remarkably insensitive to cleaning
procedures and masks already when using the $V1$ model, {apart the
case of no Bubbles/Loop-I mask and no $\ell10$ cleaning which, not
including any modeling of the Bubbles,  leaves significant residuals
which seem to bias the CCF result. Overall, this test show that the
ACF and CCF signals converge to a stable result when progressively
more aggressive cleaning procedures and latitude cuts are applied. }
For consistency with the ACF case we will consider (and show) the
CCFs computed using the same Fermi-LAT maps used for the ACF
analysis, although we did check for the robustness of all the CCFs
with respect to the different cleaning procedures and masks.

\subsubsection{Robustness to the event class type}

A further check is performed comparing the ACF and CCFs previously
obtained from {\it class 4} events maps with those computed from the
{\it class 3} and  {\it class 4} events together. In this second
case the increased contamination from residual isotropic
instrumental background is not expected to affect the CCFs although,
of course, it can make the error bars slightly larger, which is
indeed what was found. For the ACFs more effects come into play. For
example, the enhanced isotropic instrumental component is expected
to decrease the amplitude of the auto-correlation signal. However,
since the measured ACF is consistent with zero, the only effect is,
again, to slightly enlarge the error bars, which is indeed observed.
Of course, the additional instrumental background associated to {\it
class 3} events may not be completely isotropic. However, the fact
that the measured ACF is consistent with that of the  {\it class 4}
events indicates that possible deviations from isotropy are too
small to affect our analysis.

\begin{figure}
\centering \epsfig{file= 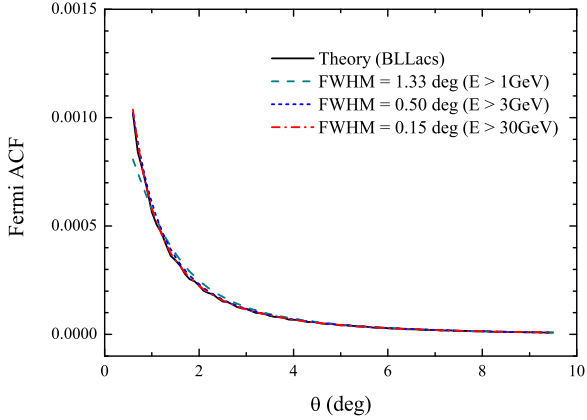, angle=0, width=0.52
\textwidth} \caption{The effect of finite PSF on the ACF of BLLacs.
\label{fig:acfw}}
\end{figure}

\subsubsection{Sensitivity to the PSF of the detector}

The angular resolution of the detector, characterized by its PSF, in
principle introduces a correlation between neighboring pixels which
may increase the auto-correlation signal on angular scales
comparable with the PSF of the detector (causing the so-called {\it
amplification bias}). The effect, however, is expected to be small
since our bin-size is significantly larger than the original
resolution of the maps. To estimate its impact on our analysis we
have first computed the window function in multipole space
associated to the PSF, given by $W_l(E)= \int d\theta
P_l(\cos(\theta)) \rm{PSF}(E, \theta) $, where $ P_l(\cos(\theta)) $
are the Legendre Polynomials and $ \rm{PSF}(E, \theta) $ is the
energy dependent PSF. We have then multiplied $W_l$  for the
expected, intrinsic correlation given by Eq.\ref{eq:angularspectrum}
and integrated back to configuration space. The result is shown in
Fig.~\ref{fig:acfw} in which we compare the expected intrinsic ACF
(continuous, black curve) to that convolved with the PSF at $E=$1
GeV, 3 GeV, and 30 GeV which has a 68\% containment radius of
$0.8^{\circ}$ (green, dot-dashed) , $0.3^{\circ}$ (blue, dotted),
$0.1^{\circ}$ (red, dashed) respectively
\citep{2009ApJ...697.1071A} . Notice also that we compared the
results with a simple Gaussian approximation for the window
function,  $W_l^2 =[\exp(-l^2\sigma_b^2/2)]^2$, where $\sigma_b$ is
the width of the beam,    finding   almost indistinguishable
results. As can be seen, only for $E\ge$1 GeV some small effects can
be appreciated, while for  $E>$3 GeV this  {\it amplification bias}
is completely negligible.

\subsubsection{Robustness to the event conversion type}

As a further robustness test we also computed the ACF of those
events labeled as \verb"front" \citep{2009ApJ...697.1071A} which are
photons converting in the front part of the detector and have a
significantly better PSF with respect to the rest of the events
which, instead, convert in the back part of the detector  where
thicker converter foils increase the chance of large-angle
scattering which deteriorates the tracking accuracy. They amount to
about half of the events detected by the Fermi-LAT detector. We
performed this test to check the robustness of the CCF with
SDSS-LRGs and 2MASS galaxies in Sections~\ref{sec:ccflrg}
and~\ref{sec:ccf2mass}, i.e. in the two cases in which we find some
features in the CCFs. Apart from some increase in the error bars due
to the halved statistics, the results  did not change significantly.

\subsection{WMAP7 ILC}
\label{sec:wmap7}

\begin{figure}
\centering \epsfig{file=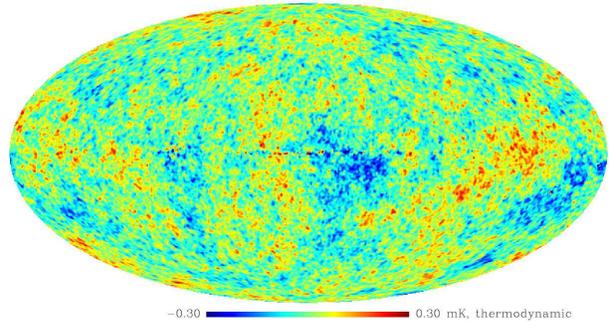, angle=0, width=0.45 \textwidth}
\caption{The 7-year WMAP ILC map in Galactic coordinates with the
resolution $N_{\rm side}=512$. \label{fig:ilc}}
\end{figure}

In order to search for the ISW effect, we cross-correlate the
Fermi-LAT 21-month EGB maps with the CMB maps derived from the
7-year WMAP data. More precisely,  we use the WMAP Internal Linear
Combination (ILC) map with $N_{\rm side} = 512$ provided by the WMAP
team \citep{2010arXiv1001.4538K}, shown in Fig. \ref{fig:ilc}. This
ILC map was already built to minimize the Galactic and other
foreground contaminations. For the WMAP map, we use the ``KQ75''
mask \citep{2010arXiv1001.4555G} corresponding roughly to the
``Kp0'' cut in the 3-year data release. In our calculations, we
downgrade it to the resolution $N_{\rm side} = 64$, to match that of
the Fermi-LAT maps, and set the weight $w_{\rm T} = 0$ for all
pixels including at least one masked high resolution pixel
\citep{2008MNRAS.386.2161R,2009JCAP...09..003X}.

\subsection{Discrete sources maps}
\label{sec:sourcemaps}

\begin{figure}
\centering \epsfig{file=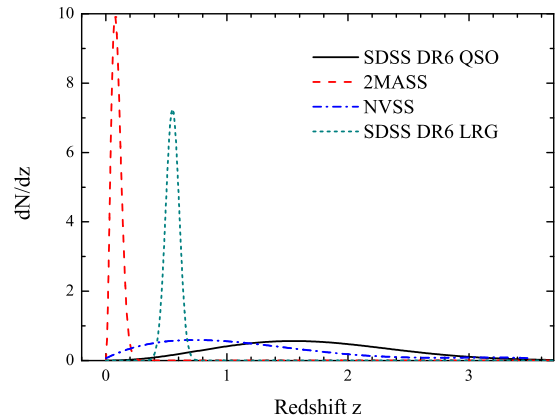, angle=0, width=0.52
\textwidth} \caption{ Normalized redshift distributions, $dN/dz$, of
the different types of objects considered for our cross-correlation
analysis. SDSS DR6 QSOs (black, continuous line), 2MASS galaxies
(red, dashed), NVSS galaxies (blue, dot-dashed) and SDSS DR6 LRGs
(cyan, short-dashed).} \label{fig:lsszdist}
\end{figure}

One of our goals is to cross-correlate the EGB maps with different
classes of sources that trace, but not necessarily coincide with,
the EGB sources. {Since all luminous objects trace, with a different
degree of bias, the same underlying distribution of matter, it makes
sense to cross correlate the EGB with the following sources: {\it
i}) Optically selected quasars, {\it ii}) luminous radio galaxies,
{\it ii}) IR-selected galaxies, and {\it iv}) LRGs. ({\it i}) and
({\it ii}) span the same, broad redshift range as the FSQRs whereas
({\it iii}) and ({\it iv}) span much narrower redshift ranges that
overlap with those of BLLacs and Starforming galaxies. } Below, we
provide some details on the different catalogs considered in our
analysis. In Fig.\ref{fig:lsszdist} we show the redshift
distribution, $dN/dz$, of the four catalogues we have considered in
this paper. All distributions are normalized to unity. 2MASS
galaxies and LRGs trace the large scale structure of the local
universe and, from  Fig.~\ref{fig:zdist1}, we see that we can expect
some cross-correlation signal only if the bulk of the EGB is
contributed by star-forming galaxies or BLLacs. On the contrary, in
the case of QSOs and NVSS galaxies, a positive cross-correlation may
be expected if EGB were preferentially contributed by a population
of high-redshift objects like FSRQs. However, the broad redshift
distribution of these objects might also allow to pick up some
cross-correlation signal provided by a population of low-redshift
$\gamma$-ray sources.

\subsubsection{SDSS DR6 QSO}
\label{sec:QSO}

\begin{figure}
\centering \epsfig{file=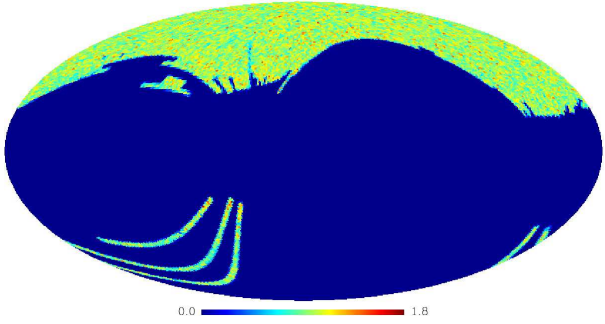, angle=0, width=0.45 \textwidth}
\caption{The number count map of SDSS DR6 quasar catalogue in
Galactic coordinates with the resolution $N_{\rm side}=64$.
\label{fig:qso}}
\end{figure}

We use the SDSS DR6 quasar catalog released by
~\citet{2009ApJS..180...67R} (hereafter DR6-QSO). This catalog
contains about $N_{\rm q}\approx10^6$ quasars with photometric
redshifts between $0.065$ and $6.075$, covering almost all of the
northern hemisphere of the Galaxy plus three narrow stripes in the
southern, for a total area of $8417~{\rm deg}^2$ ($\sim20\%$ of the
area of the whole sky). The DR6-QSO data set extends previous
similar SDSS data sets with $\sim95\%$ efficiency
\citep{2004ApJS..155..257R,2006ApJ...638..622M}. The main
differences are due to the fact that DR6-QSO probes QSOs at higher
redshift and also contains putative QSOs flagged as to have ultra
violet excess (UVX objects). We refer the reader to
\citet{2009ApJS..180...67R} for a very detailed description of the
object selection with the non-parametric Bayesian classification
kernel density estimator (NBC-KDE) algorithm.

We rely on the electronically published table that contains only
objects with the ``good'' flag with values within the range $[0,6]$.
The higher the value, the more probable for the object to be a real
QSO \citep{2009ApJS..180...67R}. We only consider the quasar candidates
selected via the UV-excess-only criteria ``uvxts=1'', i.e. objects clearly
showing a UV excess which should be a signature of a QSO spectrum.
We are left with $N_{\rm q}\approx6\times10^5$ quasars. In Fig.
\ref{fig:qso} we show the number counts map of the SDSS DR6 quasar
catalogue in Galactic coordinates.

In order to determine the mask of the actual sky coverage of the DR6
survey, we generate a random sample with a sufficiently large number
of galaxies using the DR6 database to ensure roughly uniform
sampling on the SDSS CasJobs website. Following
\citet{2009JCAP...09..003X}, besides the pixel geometry mask, we
also add the foreground mask by cutting the pixels with the g-band
Galactic extinction $A_{\rm g}\equiv3.793\times E(B-V) > 0.18$ to
account for reddening that is the main systematic effect.

The redshift distribution function $dN/dz$ of the DR6-QSO sample is
approximated by the function:
\begin{equation}
\frac{dN}{dz}(z)=\frac{\beta}{\Gamma(\frac{m+1}{\beta})}\frac{z^m}{z^{m+1}_0}
\exp\left[-\left(\frac{z}{z_0}\right)^\beta\right]~,\label{reddis}
\end{equation}
where three free parameters are $m=2.00$, $\beta=2.20$, and
$z_0=1.62$ \citep{2009JCAP...09..003X}. We choose a constant bias
$b_S=2.3$ as found by
\citet{2008PhRvD..77l3520G,2009JCAP...09..003X} to calculate the
theoretical prediction from the best-fit WMAP model adopted in this
work.

\subsubsection{2MASS}
\label{sec:2mass}

\begin{figure}
\centering \epsfig{file=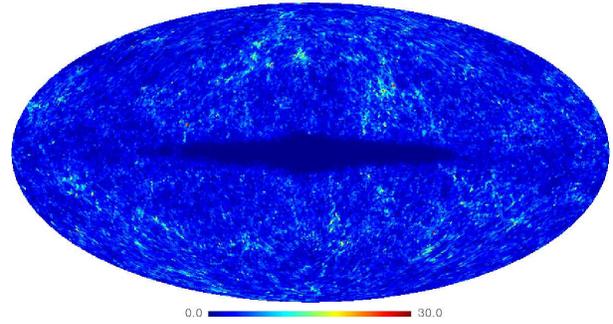, angle=0, width=0.45 \textwidth}
\caption{The number count map of 2MASS extended source catalogue in
Galactic coordinates with the resolution $N_{\rm side}=64$.
\label{fig:2mass}}
\end{figure}

We use the 2 Micron All-Sky Survey (2MASS) extended source catalogue
\citep{2000AJ....120..298J}, which contains $\sim770000$ galaxies
with mean redshift $\langle z\rangle\approx0.072$, as shown in Fig.
\ref{fig:2mass}. We select galaxies according to their $K_{\rm
s}$-band isophotal magnitude $K_{20}$, measured inside a circular
isophote with surface brightness of 20 mag/arcsec$^2$. These
magnitudes are corrected for Galactic extinction using the infrared
reddening maps: $K'_{20}=K_{20}-A_k$, where the extinction $A_{\rm
k}=0.367\times(B-V)$. In our analysis, we use the flux cut
$12.0<K'_{20}<14.0$. We only include objects with a uniform
detection threshold (${\rm use_{-}src}=1$), and remove known
artifacts (${\rm cc_{-}flag} \neq a$ and ${\rm cc_{-}flag} \neq z$).
Furthermore, we exclude areas of the sky with high reddening using
the infrared reddening maps by \citet{1998ApJ...500..525S},
discarding pixels with $A_{\rm k}> 0.05$, which leaves approximately
67\% of the sky unmasked.

In this case, the free parameters of the redshift distribution in
Eq. \ref{reddis} are $m=1.90$, $\beta=1.75$, and $z_0=0.07$
\citep{2008PhRvD..77l3520G}, while as constant bias we use $b_S=1.4$
as found by \citet{2007MNRAS.377.1085R}.

\subsubsection{NVSS}
\label{sec:nvss}

\begin{figure}
\centering \epsfig{file=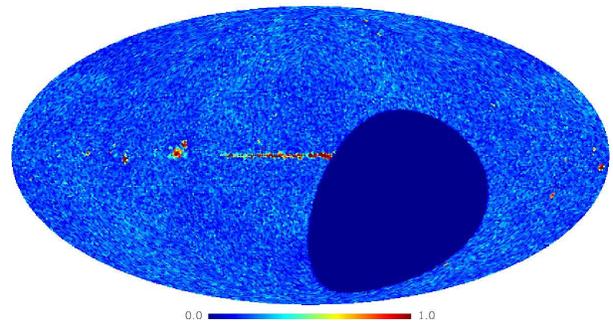, angle=0, width=0.45 \textwidth}
\caption{The number count map of NVSS radio sources in Galactic
coordinates with the resolution $N_{\rm side}=64$. \label{fig:nvss}}
\end{figure}

The NRAO VLA Sky Survey (NVSS) \citep{1998AJ....115.1693C} offers the most
extensive sky coverage ($82\%$ of the sky to a completeness limit of about 3
mJy at 1.4 GHz) and contains $1.8\times10^6$ sources. Here, we
include in our analysis only NVSS sources brighter than 10 mJy,
since the surface density distribution of fainter sources suffers
from declination-dependent fluctuations \citep{2002MNRAS.337..993B}. We also
exclude the strip at $|b| < 5^\circ$, where the catalog may be
substantially affected by Galactic emissions. The NVSS source
surface density at this threshold is 16.9 ${\rm deg}^{-2}$.

The redshift distribution at this flux limit has been recently
determined by \citet{2008MNRAS.385.1297B}. Their sample, complete to
a flux density of 7.2 mJy, comprises 110 sources with $S\geq 10$
mJy, of which 78 (71\%) have spectroscopic redshifts, 23 have
redshift estimates via the $K-z$ relation for radio sources, and 9
were not detected in the $K$-band and therefore have a lower limit
to $z$. We adopt here the smooth description of this redshift
distribution given by \citet{2010A&ARv..18....1D}:
\begin{equation}
\frac{dN}{dz}(z)=1.29 + 32.37z - 32.89z^2 + 11.13z^3 - 1.25z^4~.
\end{equation}
Here, we simply use a constant bias $b_S=1.5$ to calculate the
theoretical prediction.

\subsubsection{SDSS DR6 LRG}
\label{sec:lrg}

\begin{figure}
\centering \epsfig{file=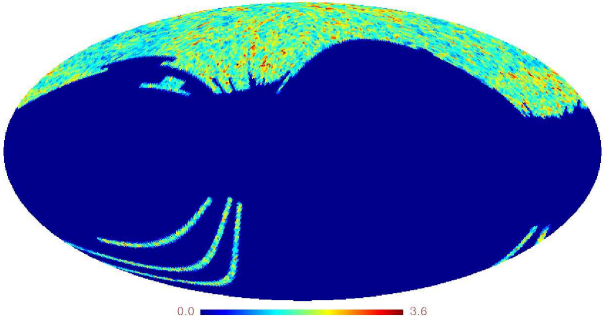, angle=0, width=0.45 \textwidth}
\caption{The number count map of MegaZ LRG catalogue in Galactic
coordinates with the resolution $N_{\rm side}=64$. \label{fig:lrg}}
\end{figure}

We use the updated MegaZ LRG sample \citep{2008arXiv0812.3831A},
which contains $\sim1.5\times 10^6$ galaxies from the SDSS DR6 in
the redshift range $0.4 < z < 0.7$ with limiting magnitude $i < 20$,
as shown in Fig. \ref{fig:lrg}. To reduce stellar contamination,
there is a variable of the MegaZ neural network estimator
$\delta_{\rm sg}$, defined such that $\delta_{\rm sg}=1$ if the
object is a galaxy, and $\delta_{\rm sg}=0$ if it is a star. For a
conservative analysis, we choose a cut $\delta_{\rm sg} > 0.2$,
which is reported to reduce stellar contamination below 2\% while
keeping 99.9\% of the galaxies. In addition to the SDSS DR6 geometry
mask, we also add two foreground masks to account for seeing
(removing pixels with median seeing in the red band larger than 1.4
arcsec) and reddening (removing pixels with median extinction in the
red band $A_{\rm r} > 0.18$) \citep{2008PhRvD..77l3520G}.

The redshift distribution function in this case is found directly
from the photometric redshifts that are given in the catalogue
\citep{2007MNRAS.375...68C}, while we set the constant bias $b_S$ to
1.8 \citep{2008PhRvD..77l3520G} when calculating the theoretical
predictions.

\section{Correlation Analysis}
\label{sec:corranalysis}

In order to calculate the ACF and the CCF all maps have been
rebinned to a resolution of $N_{\rm  side} = 64$. To estimate the
ACF of the residual Fermi-LAT maps we use the following estimator
$\hat{c}^{\rm ff}(\theta)$:
\begin{equation}
\hat{c}^{\rm
ff}(\theta)=\frac{1}{\bar{n}^2N_{\theta}}\sum_{i,j}{(n_i-\bar{n})(n_j-\bar{n})}~,
\end{equation}
where $n_{\rm i}$ is the number of $\gamma$-ray photons in each
pixel and $\bar{n}$ is the mean photon counts corresponding to the
average EGB signal. The sum runs over all the pixels with a given
angular separation. For each angular bin centered around $\theta$,
$N_{\theta}$ is the number of pixels pairs separated by an angle
within that bin. The CMB maps consist of temperature difference
maps. Therefore we replace the $n_{\rm i}$ and $\bar{n}$ with the
temperature $T_{\rm i}$ in each pixel and the average temperature
$\bar{T}$ of the CMB map. The CCF estimator (in $\mu$K) between the
Fermi-LAT map and the CMB temperature map $\hat{c}^{\rm fT}(\theta)$
reads:
\begin{equation}
\hat{c}^{\rm
fT}(\theta)=\frac{1}{\bar{n}N_{\theta}}\sum_{i,j}(T_i-\bar{T})(n_j-\bar{n})~.
\end{equation}
Moreover, we are also interested in the cross-correlation between
Fermi-LAT maps and the different maps of discrete sources. In this
case, the CCF estimator is $\hat{c}^{\rm fg}(\theta)$:
\begin{equation}
\hat{c}^{\rm
fg}(\theta)=\frac{1}{\bar{n}N_{\theta}}\sum_{i,j}{(n^{\rm
f}_i-\bar{n}^{\rm f})}\frac{(n^{\rm g}_j-\bar{n}^{\rm
g})}{\bar{n}^{\rm g}}~.
\end{equation}
Since we are using a resolution of $N_{\rm side} = 64$, for which
the pixel size is $0.92^{\circ}$, in our calculation we use $N_{\rm
bin} = 5$ linearly spaced angular bins in the range
$1^\circ\leq\theta\leq9^\circ$.

To compute the errors, we estimate the covariance matrix of the ACF
(CCF) data points using jackknife resampling method
\citep{2002ApJ...579...48S}. This method divides the data into $M$
patches, then $M$ subsamples are created by neglecting each patch in
turn. These patches have roughly equal area. In practice, we firstly
list the whole set of pixels covered by the survey, and then divide
them into $M = 30$ patches that do not have very similar shape, but
have roughly equal area (i.e. equal number of pixels). The
covariance estimator is:
\begin{equation}
C_{ij}=\frac{M-1}{M}\sum^M_{k=1}\left[\hat{C}_k(\theta_i)-\bar{C}(\theta_i)\right]\left[\hat{C}_k(\theta_j)-\bar{C}(\theta_j)\right]~,
\end{equation}
where $\hat{C}_k(\theta_i)$ are the observed ACF (CCF) of the $M$
subsamples in the $i$-th angular bin and $\bar{C}(\theta_{\rm i})$
are the mean ACF (CCF) over $M$ realizations. The diagonal part of
these matrices gives the variance of the ACF (CCF) in each bin
$C^k_{ii}=\sigma^2_{\rm i}$, while the off-diagonal part represents
the covariance between the angular bins. We also change the number
of patches $M$ and verify that the covariance matrix is stable.

\section{Results}
\label{sec:results}

In this section we compute the ACF and the CCF of the different map
combinations and compare the results  with model predictions. The
ACF signal could constrain the nature of the sources that contribute
to the EGB. The presence of the ISW effect would quantify how well
the EGB sources trace the underlying mass distribution and reveal
the presence of a cosmological constant term. Finally, the strength
of the CCF signal would indicate how closely a given class of
objects trace the sources of the EGB. {For each correlation analysis
we will show measured quantities and theoretical predictions. The
comparison between the two is only performed in configuration space,
i.e. we over-plot the theoretical ACF and CCF (continuous curves)
with the data points with error bars. In addition, we also will show
mode predictions in Legendre space, i.e. through the expected
angular power spectra, since this is the direct prediction of the
model, and  model to model comparison is easier in Legendre space.
All theoretical predictions use Eq.\ref{eq:angularspectrum} to model
the correlation between the $\gamma$-ray signal supposedly
contributed by a single type of objects (FSRQs, BLLacs or Star
forming galaxies) and a second fluctuation fields (the $\gamma$-ray
background for the ACF, the CMB for the ISW signal and the discrete
source catalogs for the cross-correlation analyses). }

\subsection{Auto-correlation analysis}
\label{sec:acf}

\begin{figure}
\centering \epsfig{file=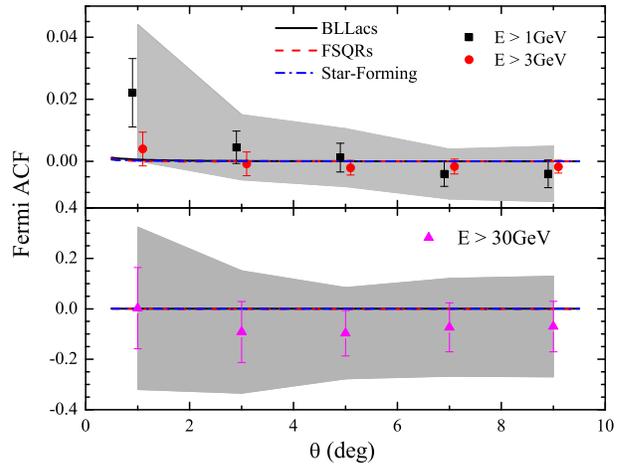, angle=0, width=0.52
\textwidth} \caption{ACFs estimated from the 21-month Fermi-LAT EGB
map for  $|b| > 20^{\circ}$ in three energy bands: $E>1$ GeV, $E>3$
GeV (upper panel) and $E>30$ GeV (below panel). Model predictions
for different types of sources are represented by continuous curves:
FSRQs (black, continuous), BLLacs (red, dashed) star-forming
galaxies (blue, dot-dashed). The gray band indicates the systematic
uncertainty coming from the foreground cleaning procedure estimated
to be approximately equal to the 1$\sigma$ statistical errors (see
text). In the top panel only the band for the 1 GeV case is shown.
\label{fig:acf_3}}
\end{figure}

In Fig.\ref{fig:acf_3} we show the ACF of the residual maps obtained
using the $V2$ Galactic model with $\ell10$ cleaning, Bubbles/Loop-I
masking and a Galactic cut $|b| = 20^{\circ}$. Different symbols
indicate  different energy cuts: black square is for $E>1$ GeV, red
dot is for $E>3$ GeV and magenta triangle is for $E>30$ GeV. Results
for the $E>30$ GeV case are shown in separate panel since the error
bars are significantly larger. 1$\sigma$ error bars were computed
using the jackknife procedure. The three continuous curves in this
plot and in all plots of the cross-correlation functions discussed
in the following sections represent the theoretical predictions from
Section~\ref{sec:correlation} obtained assuming that each source
class (BLLacs, FSRQs or Star Forming galaxies), specified by their
luminosity density distribution $\rho_{\gamma}(z)$, contribute to a
fraction of the EGB, $f_j$, listed  in Table~\ref{tab:tab1}.

At small angular separations ($\theta<2^{\circ}$) the
auto-correlation signal is consistent with zero with all energy cuts
but $E>1$ GeV. Theoretical models do predict a weak auto-correlation
signal at  these angular separations that can be regarded as the
typical angular size of the $\gamma$-ray emitting element. However,
the predicted auto-correlation is much weaker than the measured one,
at a level that would be indistinguishable from zero with the
current uncertainties. Although a correlation signal at this level
would still be possible from a  contribution of unresolved point
sources, we find that the signal is not very robust to the different
cleaning methods and progressively disappears when we apply larger
$|b|$ cuts. We therefore regard this as a spurious feature of the
ACF  at low energies where the Galactic contribution is stronger and
its subtraction more prone to systematic errors. At larger
separations ($\theta>4^{\circ}$) the ACF is slightly negative. This
is a spurious features induced by the $l10$ cleaning procedure as we
have shown the Monte Carlo analysis presented in Section~\ref
{sec:montecarlo}. The corrected signal would be consistent with
zero. {From the Montecarlo we estimate that this systematic offset
is at the level of  the  1$\sigma$ statistical error. We thus show
in the plot also a systematic uncertainty band obtained doubling the
1$\sigma$ statistical uncertainties. }

\begin{figure}
\centering \epsfig{file=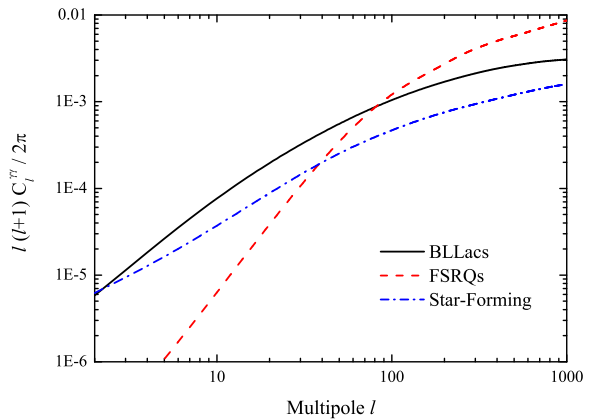, angle=0, width=0.52
\textwidth} \caption{Angular auto-power spectra for different EGB
source models. Different line-styles characterize different models.
Black, continuous curve: FSRQs; red, dashed: BLLacs; blue,
dot-dashed: star-forming galaxies. \label{fig:acf_cl}}
\end{figure}

To highlight the differences among the different source classes we
plot the angular auto-power spectra in Fig.~\ref{fig:acf_cl}. BLLacs
and star-forming galaxies have similar spectra, with more power on
large scales than the FSRQ model. The situation is reversed at small
scales. This difference simply reflects the fact that in the first
two cases the $\gamma$-ray emission peaks at moderate redshifts
$z=[0.5,1]$ while for FSRQs the bulk of the $\gamma$-ray signal is
produced at $z>2$, as shown in Fig.~\ref{fig:zdist1}. Since all
models trace the same mass density field, i.e. assume the same
$P(k)$, the power shift in  Fig.~\ref{fig:acf_3} reflects the fact
that the same physical scale is preferentially seen at different
angles in the different models: large angles (small $\ell$) for
BLLacs and star-forming galaxies that typically sample the universe
at low redshifts; small angles (large $\ell$) for FSRQs  that
preferentially samples the universe at high redshifts. The larger
amplitude of the spectrum for BLLacs to the spectrum of star-forming
galaxies simply reflects the different bias factors of the two
classes of objects.

\subsection {ISW}
\label{sec:isw}

\begin{figure}
\centering \epsfig{file=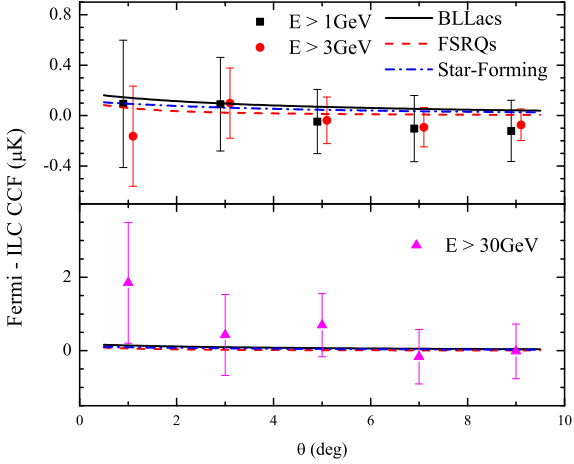, angle=0, width=0.52
\textwidth} \caption{CCFs estimated from the WMAP7 ILC map and the
21-month Fermi-LAT EGB map with $|b| > 20^{\circ}$ in three energy
bands. The three symbols refer to 3 energy cuts $E>1$ GeV, $E>3$ GeV
(upper panel) and $E>30$ GeV (below panel). Model predictions for
different types of sources are represented by continuous curves:
FSRQs (black, continuous), BLLacs (red, dashed), star-forming
galaxies (blue, dot-dashed). \label{fig:fermi3_ccf_ilc_Rescaling}}
\end{figure}

In Fig.~\ref{fig:fermi3_ccf_ilc_Rescaling} we show the CCF of the
EGB with the CMB, i.e. the ISW signal. The symbols refer to the CCF
estimated from the 21-month Fermi-LAT maps  and the WMP7 ILC map and
the continuous lines represent model predictions. Different symbols
indicate different energy cuts.

{We note that the expected  CCF signal is positive out to large
angular separations and is only a factor 3-4 smaller than the width
of the error bars. This has to be compared with the ACF case where
the expected signal is, instead, more than a factor of 10 smaller
than the the experimental error bars. This result, which reflects
the better statistics of the CMB maps, indicates that the goal of
detecting the ISW signature in the EGB is not unrealistic. However,
with the limited statistics of the 21-month data the ISW signal is
consistent with zero with all energy cuts. This (null) result is
robust to different cleaning methods and latitude cuts.}

\begin{figure}
\centering \epsfig{file=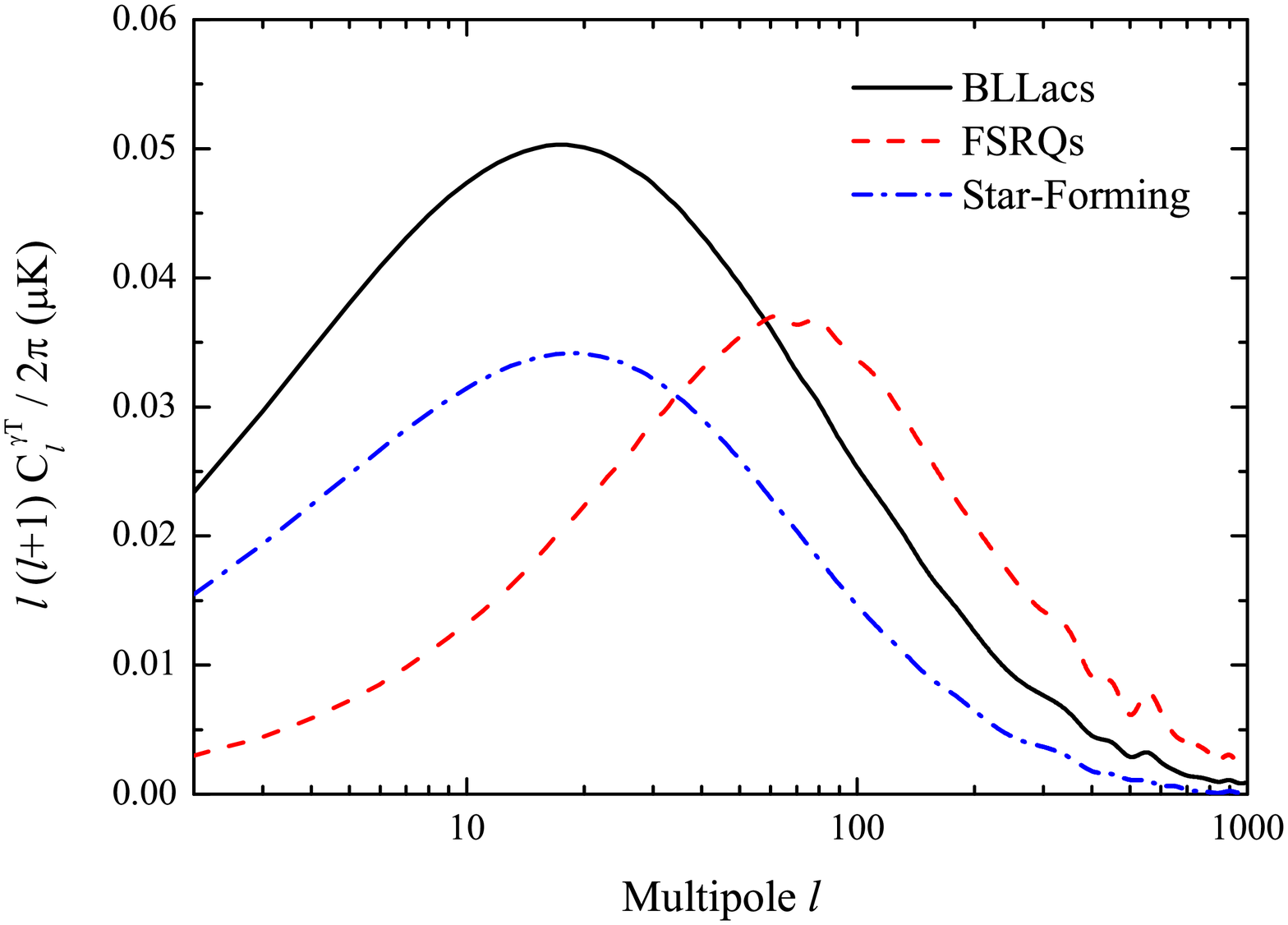, angle=0, width=0.52
\textwidth} \caption{Angular cross-power spectra of the EGB with the
CMB. Different line-styles characterize different models. Black,
continuous curve: FSRQs. Red, dashed: BLLacs. Blue, dot-dashed:
star-forming galaxies. \label{fig:fermi_ilc_ccf_cl}}
\end{figure}

From a theoretical point of view we notice that the ISW signal is
expected to be larger in the BLLacs case than in the FSRQs one. This
is because the former sample is rather local $z\le 1$ and probes an
epoch in which the cosmological constant drives the accelerated
expansion in which the late ISW effect sets in. FSRQs, instead,
preferentially sample a high redshift, matter-dominated, flat
universe from which we expect only a very weak ISW signature.  Once
again, the difference between the BLLacs and star-forming galaxies
models reflects the different biases of the two populations.  The
difference among the models is best seen through their angular
cross-spectra, shown in Fig.~\ref{fig:fermi_ilc_ccf_cl}.  As for the
auto-power case, we notice that the FSRQs cross-power peaks at
smaller angles than that of the other models, reflecting the
different redshift sampled by the different classes of objects. The
smaller power of the FSRQs model reflects the intrinsic weakness of
the ISW signal, as anticipated. The drop at $l\sim 100$ is due to
the late-time (low $\ell$) nature of the ISW effect.

\subsection{Cross-correlation with QSOs}
\label{sec:ccfqso}

\begin{figure}
%\vspace{-1cm}
%\centering \epsfig{file=ccf_qso_1gev_20deg.eps, angle=0, width=0.49 \textwidth}
%\vspace{-1cm}
%\centering \epsfig{file=ccf_qso_3gev_20deg.eps, angle=0, width=0.49 \textwidth}
%\vspace{-1cm}
%\centering \epsfig{file=ccf_qso_30gev_20deg.eps, angle=0, width=0.49 \textwidth}
%\vspace{-0.5cm}
\centering \epsfig{file=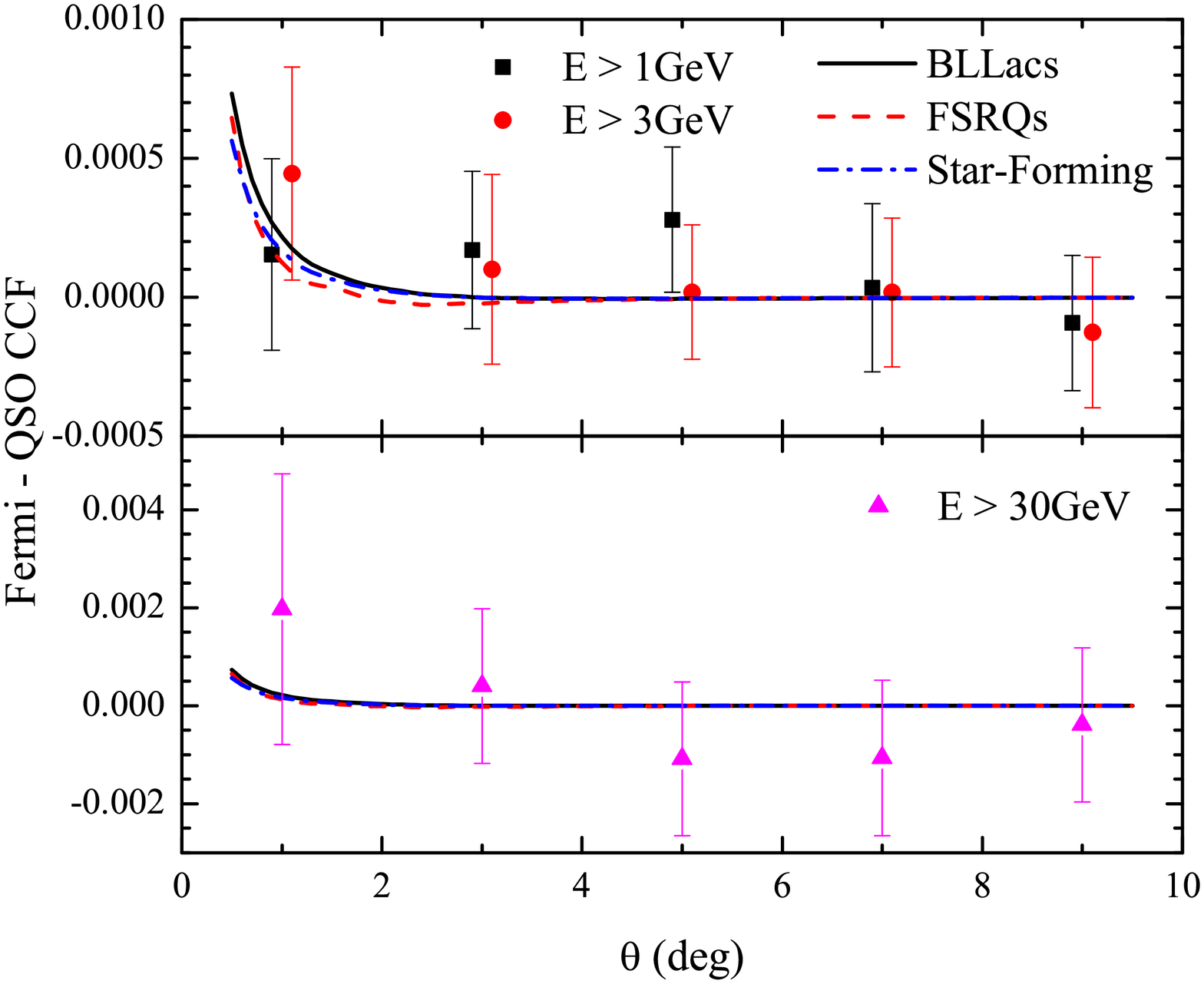, angle=0, width=0.52
\textwidth} \caption{CCFs estimated from the SDSS DR6 QSO map and
the 21-month Fermi-LAT EGB map with $|b| > 20^{\circ}$ in three
energy bands. The three symbols refer to 3 energy cuts $E>1$ GeV,
$E>3$ GeV (upper panel) and $E>30$ GeV (below panel). Model
predictions for different types of sources are represented by
continuous curves: FSRQs (black, continuous), BLLacs (red, dashed)
star-forming galaxies (blue, dot-dashed).
\label{fig:qso_ccf_fermi_Rescaling}}
\end{figure}

\begin{figure}
\centering \epsfig{file=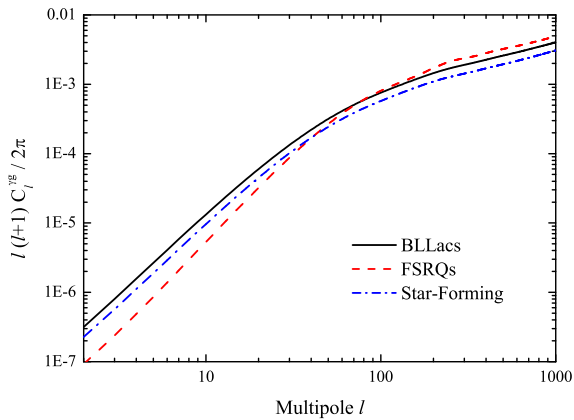, angle=0, width=0.52
\textwidth} \caption{Angular cross-power spectra of the EGB with the
SDSS-DR6 QSOs. Different line-styles characterize different models.
Black, continuous curve: FSRQs. Red, dashed: BLLacs. Blue,
dot-dashed: star-forming galaxies. \label{fig:fermi_qso_ccf_cl}}
\end{figure}

In Fig.~\ref{fig:qso_ccf_fermi_Rescaling} we plot the estimated CCF
between the Fermi-LAT EGB and the SDSS-DR6 QSOs maps compared with
model predictions. Symbols and line-styles are the same as in
Fig.~\ref{fig:fermi3_ccf_ilc_Rescaling}. The expected CCF amplitude
is rather weak. This is due to the quite sharp, high redshift peaks
of the quasars' $dN/dz$ which preferentially picks up the
correlation signal from high redshifts. With a large $z$ value, the
$(1+z)^{-\Gamma} H^{-1}(z)$ dimming term in Eq.~\ref{eq:autocorr}
weakens the correlation signal. The fact that the correlation signal
comes from large redshift irrespective of the EGB model explored, is
more clearly seen in Fig.~\ref{fig:fermi_qso_ccf_cl} which shows
that the angular power has indeed been driven toward small angular
scales. {We notice here that it is still physical to consider, for
example, the star-forming galaxies model in relation to the
cross-correlation with QSOs. Indeed, a cross-correlation analysis
alone cannot unambiguously identify EGB sources since all of them
trace, with different biases, the underlying Dark Matter field. So,
there may be a positive cross-correlation with the SDSS-DR6 QSOs
catalog, which hardly consists of star forming galaxies,  even if
the underlying sources of the EGB are star-forming galaxies,  since
both of them trace the  the Dark Matter field.}

All CCFs are consistent with zero. The weak ($\sim 1 \sigma$)
positive correlation in the innermost bin for the $E>3$ GeV case
(which would be consistent with theoretical expectations),
disappears for the different choices of the cleaning method and/or
when increasing the $|b|$ cut.

Note that, although the SDSS DR6 QSO catalog has very high
efficiency in the selection algorithm, stars are point-like sources
that inevitably contaminate the catalog. To compute the stellar
contamination, we extract a large number ($\sim 8\times10^4$) of
stars from the SDSS DR6 survey in the magnitude range $16.9 < g <
17.1$ using the CasJobs \footnote{http://casjobs.sdss.org/dr7/en/}
website and compute the CCF between stars and the Fermi maps
$\hat{c}^{\rm fs}(\theta)$. We have carefully checked that the
contribution from contaminating stars to the CCF between QSOs and
Fermi-LAT maps is fully consistent with zero and can safely be
neglected.

\subsection{Cross-correlation with NVSS galaxies}
\label{sec:ccfnvss}

\begin{figure}
\centering \epsfig{file=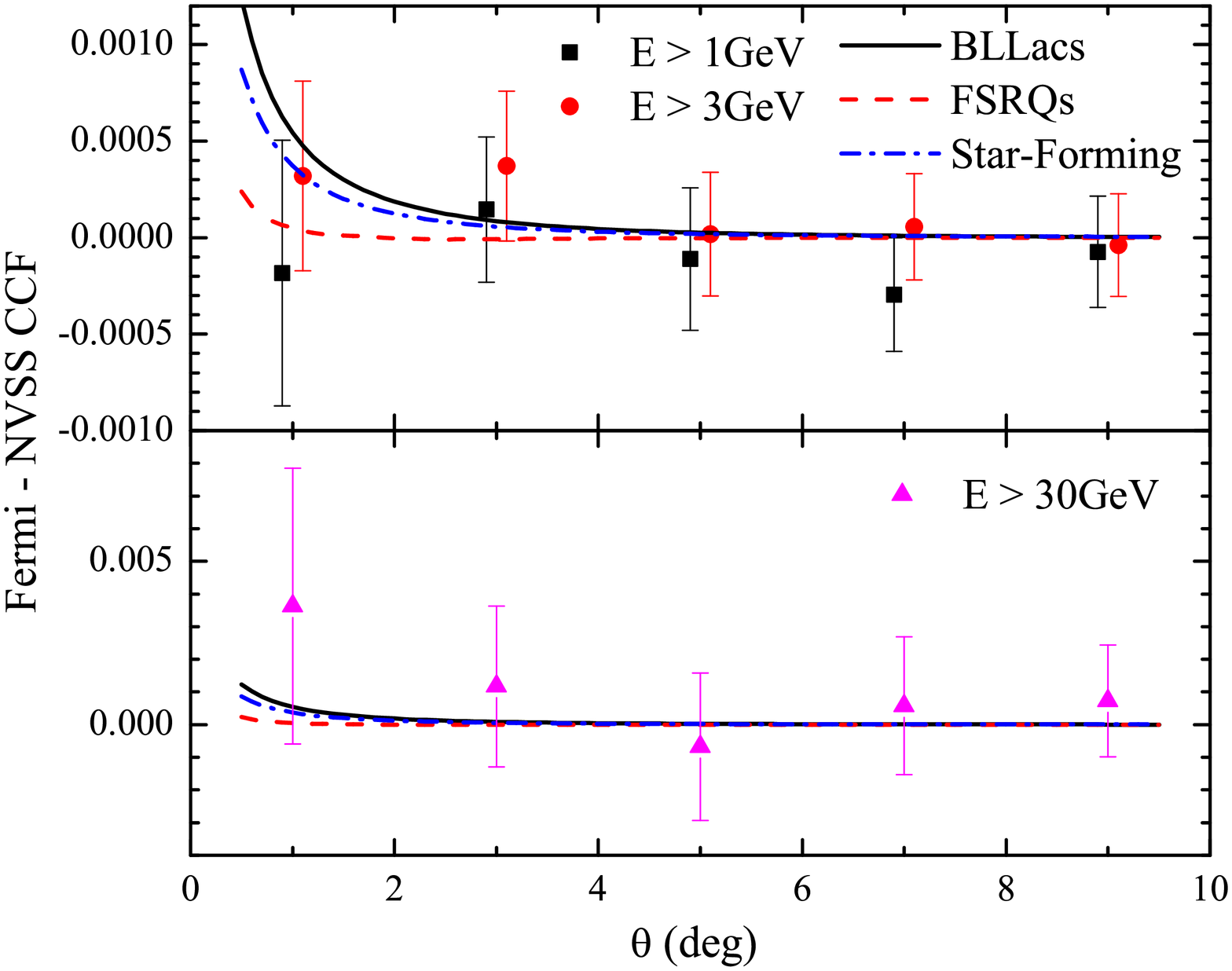, angle=0, width=0.52
\textwidth}
%\vspace{-1cm}
%\epsfig{file=ccf_nvss_3gev.eps, angle=0, width=0.49 \textwidth}
%\vspace{-0.5cm}
%\epsfig{file=ccf_nvss_30gev.eps, angle=0, width=0.49 \textwidth}
%\vspace{-0.5cm}
\caption{CCFs estimated from the NVSS galaxies map and the 21-month
Fermi-LAT EGB map with $|b| > 20^{\circ}$ in three energy bands. The
three symbols refer to 3 energy cuts $E>1$ GeV, $E>3$ GeV (upper
panel) and $E>30$ GeV (below panel). Model predictions for different
types of sources are represented by continuous curves: FSRQs (black,
continuous), BLLacs (red, dashed) star-forming galaxies (blue,
dot-dashed). \label{fig:nvss_ccf_fermi_Rescaling}}
\end{figure}

\begin{figure}
\centering \vspace{-1cm} \epsfig{file=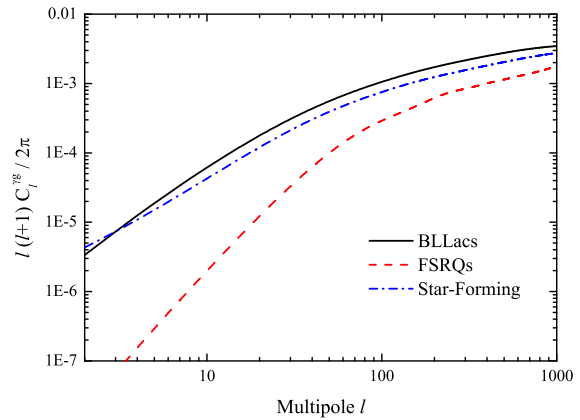,
angle=0, width=0.52 \textwidth} \caption{Angular cross-power spectra
of the EGB with the NVSS galaxies. Different line-styles
characterize different models. Black, continuous curve: FSRQs. Red,
dashed: BLLacs. Blue, dot-dashed: star-forming galaxies.
\label{fig:fermi_nvss_ccf_cl}}
\end{figure}

The $dN/dz$ of NVSS galaxies peaks at lower redshifts than the QSO
ones. For this reason the expected CCF signal is larger than for
QSOs (see Fig.~\ref{fig:nvss_ccf_fermi_Rescaling}) and the
cross-power shifts at larger angles (see
Fig.~\ref{fig:fermi_nvss_ccf_cl}). FSRQs have a weaker
cross-correlation signal (and cross-power) because there is little
overlap between the NVSS $dN/dz$ and the redshift distribution of
the $\gamma$-ray emission signal (see Fig.~\ref{fig:zdist1}).

The estimated CCFs are consistent with zero and with theoretical
predictions, despite the fact that all Fermi sources, including
normal galaxies, are also NVSS sources. At the moment, this results
is not worrisome given that the expected signal still fits within
the large error bars. It will be indeed interesting to see if it
persists with larger statistics.

\subsection{Cross-correlation with LRGs galaxies}
\label{sec:ccflrg}

\begin{figure}
\centering
%\vspace{-1cm}
%\epsfig{file=ccf_lrg_1gev_20deg.eps, angle=0, width=0.49 \textwidth}
%\vspace{-1cm}
%\epsfig{file=ccf_lrg_3gev.eps, angle=0, width=0.49 \textwidth}
%\vspace{-0.5cm}
\epsfig{file=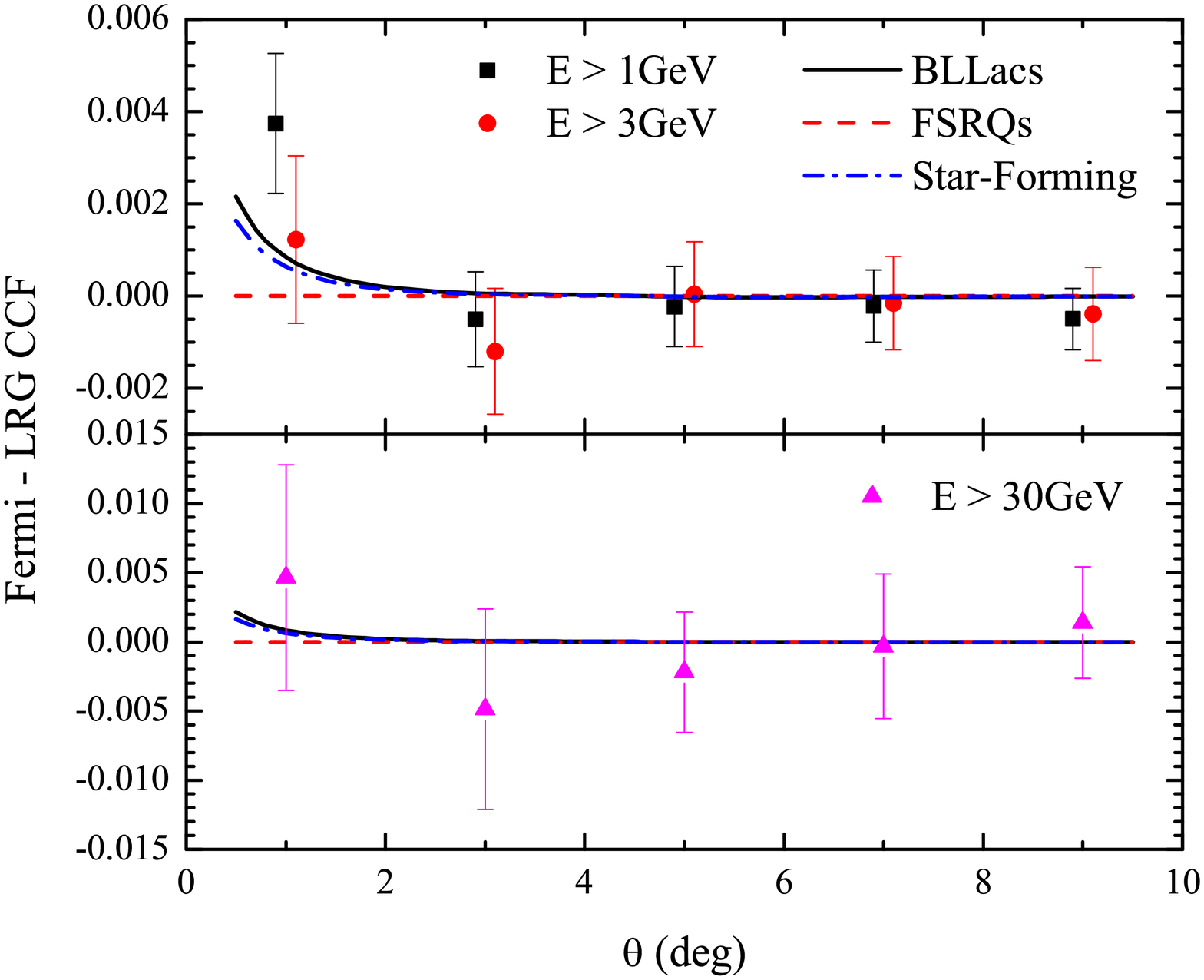, angle=0, width=0.52 \textwidth}
\caption{CCFs estimated from the LRGs map and the 21-month Fermi-LAT
EGB map with $|b| > 20^{\circ}$ in three energy bands. The three
symbols refer to 3 energy cuts $E>1$ GeV, $E>3$ GeV (upper panel)
and $E>30$ GeV (below panel). Model predictions for different types
of sources are represented by continuous curves: FSRQs (black,
continuous), BBLacs (red, dashed), star-forming galaxies (blue,
dot-dashed). \label{fig:lrg_ccf_fermi_Rescaling}}
\end{figure}

\begin{figure}
\centering \epsfig{file=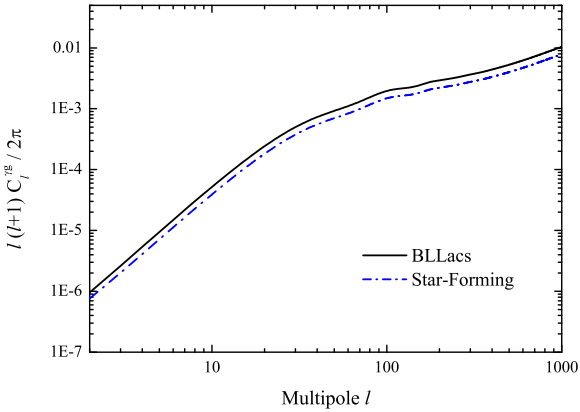, angle=0, width=0.52
\textwidth} \caption{Angular cross-power spectra of the EGB with the
LRGs. Different line-styles characterize different models. Black,
continuous curve: BLLacs. Blue, dot-dashed: star-forming galaxies.
\label{fig:fermi_lrg_ccf_cl}}
\end{figure}

Fig.~\ref{fig:lrg_ccf_fermi_Rescaling} shows that the expected
amplitude of the CCF with the LRGs for BLLacs and star-forming
galaxies is larger than that of NVSS galaxies because the LRGs
distribution peaks at similar scales but is much sharper. As a
consequence the peak of the angular cross-power  for BLLacs and
star-forming galaxies is quite similar to the NVSS one with some
more power at small angles (see Fig.~\ref{fig:fermi_lrg_ccf_cl}).
There is no curve for the FSRQs case since this model predicts zero
cross-correlation signal because there is no  overlap between the
$dN/dz$ distribution of the LRGs and the predicted redshift
distribution of $\gamma$-ray signal.

The measured CCFs  is consistent with zero for  $E>3$ and $E>30$
GeV. With the lowest energy cut, $E>1$ GeV, we detect a positive
correlation signal at $\theta<2^{\circ}$ at $\sim 2 \sigma$
confidence. This signal is remarkably robust to cleaning procedures
and Galactic cuts. This feature is also robust to the choice of the
$\gamma$-ray events since it is also present when we only consider
the so-called \verb"front"  $\gamma$-ray photons which have a
significantly better PSF.

Theoretical predictions agree with this signal at the $\sim
1.5\sigma$ level. The fact that is only seen at low energies  may
indicates that the sources that contribute to the EGB at low
energies are brighter than expected in our power-law model, i.e. the
bias of the EGB sources ($b_\gamma$ in Eq.\ref{eq:autocorr}) is
larger than expected. More intriguingly, this may indicate that
there is a transition in energy in the sources contributing to the
EGB from e.g. galaxies or BLLacs at low energy (which
cross-correlate with LRGs) to FSRQs at high energies (which do not).

A further alternative is that this signal may come from the
cross-correlation with the $\gamma$-rays contributed by sources
coincident with LRGs but that are still too faint to be detected.
Luckily, more statistics and better understanding of the diffuse
foregrounds will help in the near future to better characterize this
feature (for example using finer angular bins or lower energy
photons) and check the validity of the previous hypotheses.

\subsection{Cross-correlation with 2MASS galaxies}
\label{sec:ccf2mass}

\begin{figure}
%\vspace{-1cm}
%\centering \epsfig{file=ccf_mass_1gev_20deg.eps, angle=0, width=0.49 \textwidth}
%\vspace{-1cm}
%\centering \epsfig{file=ccf_2mass_3gev.eps, angle=0, width=0.49 \textwidth}
%\vspace{-0.5cm}
\centering \epsfig{file=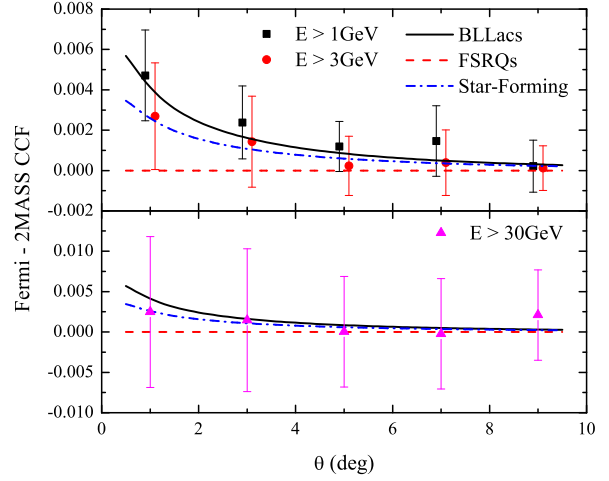, angle=0, width=0.52
\textwidth} \caption{CCFs estimated from the 2MASS map and the
21-month Fermi-LAT EGB map with $|b| > 20^{\circ}$ in three energy
bands. The three symbols refer to 3 energy cuts $E>1$ GeV, $E>3$ GeV
(upper panel) and $E>30$ GeV (below panel). Model predictions for
different types of sources are represented by continuous curves:
FSRQs (black, continuous), BLLacs (red, dashed) star-forming
galaxies (blue, dot-dashed). \label{fig:2mass_ccf_fermi_Rescaling}}
\end{figure}

\begin{figure}
\centering \epsfig{file=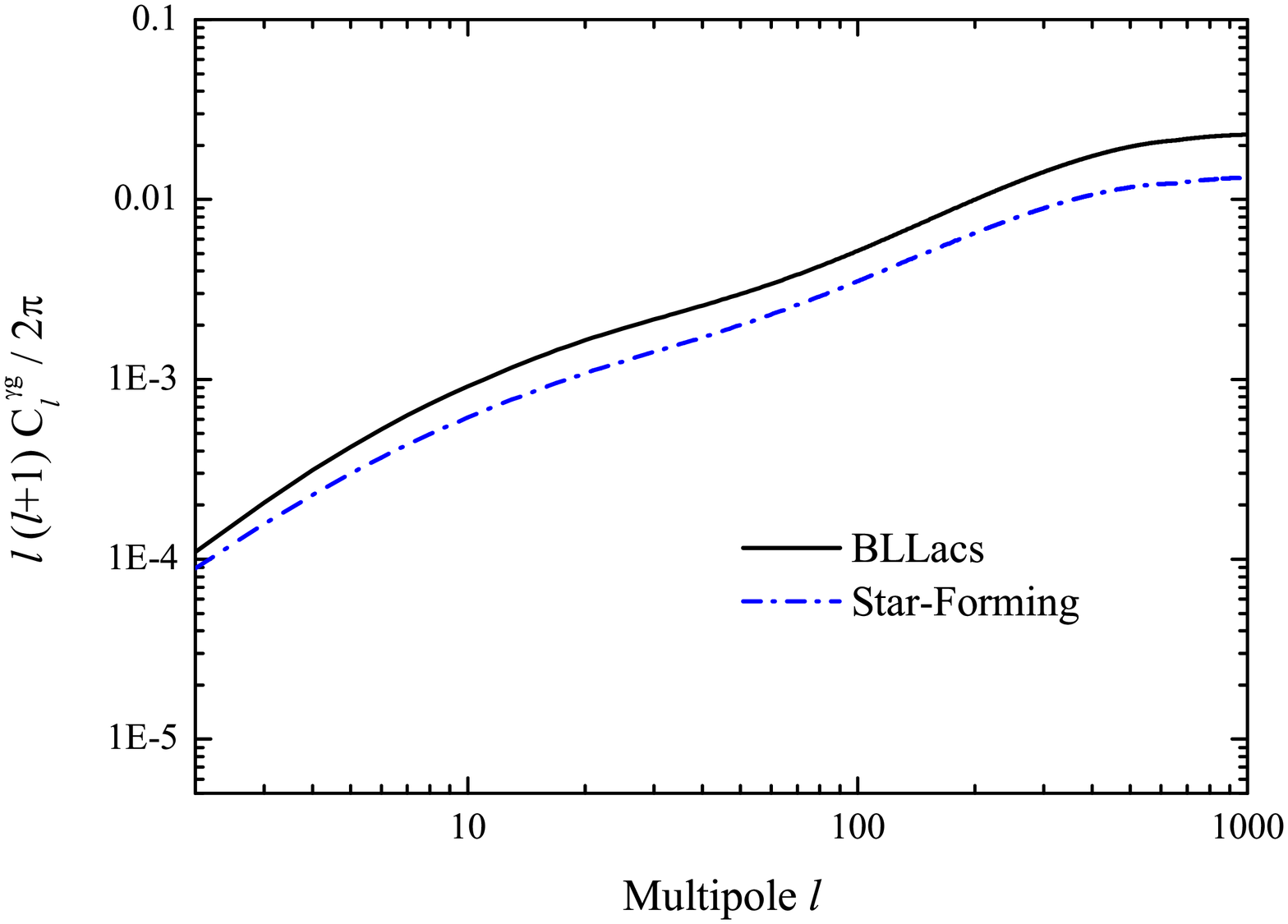, angle=0, width=0.52
\textwidth} \caption{Angular cross-power spectra of the EGB with the
2MASS galaxies. Different line-styles characterize different models.
Black, continuous curve: BBLacs. Blue, dot-dashed: star-forming
galaxies. \label{fig:fermi_2mass_ccf_cl}}
\end{figure}

The result of the cross-correlation between Fermi-LAT EGB maps and
2MASS catalog (that represents the most local of our samples)
confirms the trend of the other cross-correlation analyses: increase
of the expected CCF amplitude predicted by the models of BLLacs and
star-forming galaxies (see
Fig.~\ref{fig:2mass_ccf_fermi_Rescaling}), angular power that shifts
toward larger angles (see Fig.~\ref{fig:fermi_2mass_ccf_cl}) and
zero correlation expected for an EGB solely contributed by FSRQs.
{We note that, as expected, the theoretical angular  cross-spectrum
is in good agreement with the one computed by
\citet{2009MNRAS.400.2122A}. The small differences likely arise from
the fact that in our estimate we did not account for the angular
resolution of the instrument and did not filter the angular power
spectrum accordingly.}

The measured cross-correlation signal is consistent with zero at all
but small angular separations and for $E>1$ GeV, where we detect a
hint of positive correlation. However, the reality of this
correlation signal is questionable for two reasons. On one side, we
found that this signal is rather sensitive to the cleaning procedure
and to the Galactic mask adopted. Also in the LRGs case we made a
further check using \verb"front" events only, but the sensitivity to
data cleaning technique still persists. On the other side, this
signal could be related to some possible systematic errors in the
treatment of the 2MASS catalogue which has been advocated to settle
some controversy in the ISW detection \citep{2010MNRAS.406....2F}.

%\label{fig:2mass_ccf_fermi_Rescaling}}
%\end{figure}

\section{analysis}
\label{sec:analysis}

\begin{table}
%[ht]
\begin{center}
\caption{$\chi^2$ test of the observed data against the $\Lambda$CDM
and SCDM models. \label{tab:tab2}}
\begin{tabular}{|c|cc|cc|}
\hline\hline Sources & \multicolumn{2}{c}{Current Data ($\chi^2$)} &
\multicolumn{2}{c}{Forecast ($\Delta\chi^2$)}\\
&$\Lambda$CDM&SCDM&Five years&Ten years\\
\hline

FSRQs&19.94&20.33&0.7&1.3\\
BLLacs&19.21&19.45&0.6&1.2\\
Star-form.&20.80& 21.50&1.8&3.5\\

\hline
\end{tabular}
\end{center}
\end{table}

In this section we quantify the capability of the correlation
analyses presented in the previous sections to constrain the
cosmological model and the nature of the sources that contribute to
the EGB using the observed ACFs and CCFs. For this purpose we only
consider $E>3$ GeV maps that allowed to obtain robust results (in
contrast with the $E>1$ GeV case) with reasonable errorbars (in
contrast with the $E>30$ GeV case).

{To do this we have computed the following $\chi^2$ function:
%\begin{equation}
%\chi^2=\sum_{\gamma,k}
%{\cal G}_{\gamma,k}^{-1}
%\left[
%\sum_{i,j} { C_{i,j}^{-1}
%(d_i^{\gamma,k}-t_i^{\gamma,k})
%(d_j^{\gamma,k}-t_j^{\gamma,k})}
%\right]_{\gamma,k}
%\label{eq:chi2}
%\end{equation}
%
\begin{equation}
\chi^2= \sum_{i,j,k,l} { C_{i,j,k,l}^{-1}
(d_i^{\gamma,k}-t_i^{\gamma,k}) (d_j^{\gamma,l}-t_j^{\gamma,l})}~,
\label{eq:chi2}
\end{equation}
where $d^{\gamma,k}$ represents the measured CCF between the diffuse
EGB and the catalog $k$ (coinciding with the ACFs when $k = \gamma$)
and $t^{\gamma,k}$ is the theoretical prediction. The $i,j$ indexes
run  over the 5 angular bins of each of the various  ACF/CCFs,
while the $k,l$ indexes   run over the different  ACF/CCFs.
$C_{i,j,k,l}$ indicates the general covariance matrix  obtained from
the jackknife resampling, and  contains the correlation between
different angular bins as well as between  the various ACF/CCFs
relative to different catalogues. In practice, this $\chi^2$
statistics compares the measured ACF and CCFs presented in
Section~\ref{sec:results} for our 3 model predictions (BLLacs, FSRQs
and star-forming galaxies), considering all angular separations
$\theta \le 10^{\circ}$ and taking into account the covariance among
the different ACF/CCF estimates. }

The purpose of this comparison is twofold. First, we want to check
whether  this analysis is sensitive to the presence of a
cosmological constant (mainly through the EGB - CMB comparison).
Second, within a given cosmological framework, we want to check our
ability of discriminating among competing models for the EGB. To
answer these questions we have performed two different tests.

The results of the first test are summarized in
Table~\ref{tab:tab2}. Columns 2 and 3 list the $\chi^2$ values
obtained when we consider the ``concordance" $\Lambda$CDM model
adopted in  this paper (second column) and  the $\chi^2$ values
obtained when we consider a CDM Einstein-de Sitter model with
$\Omega_{\rm m}=1.0$ (SCDM, third column). The different rows refer
to the three EGB models considered (column 1). The  $\chi^2$ values
obtained for the two cosmological models are rather small. This
similarity reflects the insensitivity of our correlation analysis to
the presence of a cosmological constant term. The situation will
improve significantly with the future Fermi-LAT data. In columns 4
and 5 we list the increase of the $\chi^2$, relative to the values
in column 2, expected after $t_{\rm obs}=5$ and $t_{\rm obs}=10$
years of observations, respectively. This forecast has been obtained
by assuming Poisson errors i.e. by scaling the 21-month errorbars by
$\sqrt t_{\rm obs}$. One can see that with $t_{\rm obs}=10$  years
one expects to discriminate a $\Lambda$CDM model form a  SCDM
scenario at about $2\,\sigma$ confidence level if the EGB were
mainly contributed by star-forming galaxies. This estimate assumes
Gaussian statistics and refers to the case of one free parameter
only: the value of $\Omega_{\rm m}$. In fact the situation is likely
to be more favorable since future data will also allow to improve
the Galactic foreground model. In this way we will be able to extend
the correlation analyses to lower energy bands. In addition, one can
improve the effectiveness of the $\chi^2$ statistics by carefully
selecting the range of angular scales to be considered or by
restricting the analysis to a few CCFs, among which the CCF with the
CMB will play a crucial role since, as we have pointed out, the ISW
signal is very sensitive to the underlying cosmological model.

\begin{figure*}
\centering

\epsfig{file=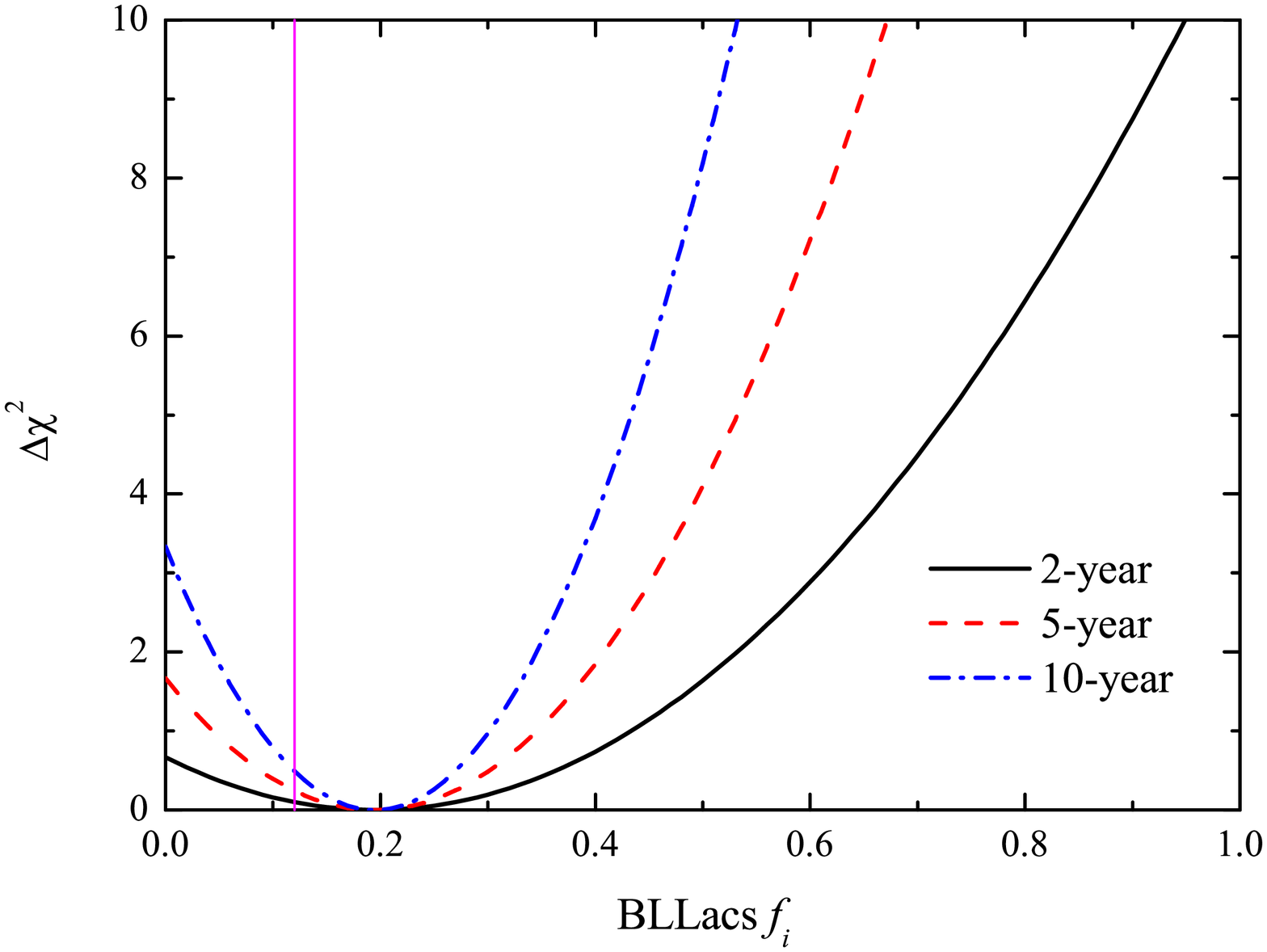, angle=0, width=0.33 \textwidth}
\epsfig{file=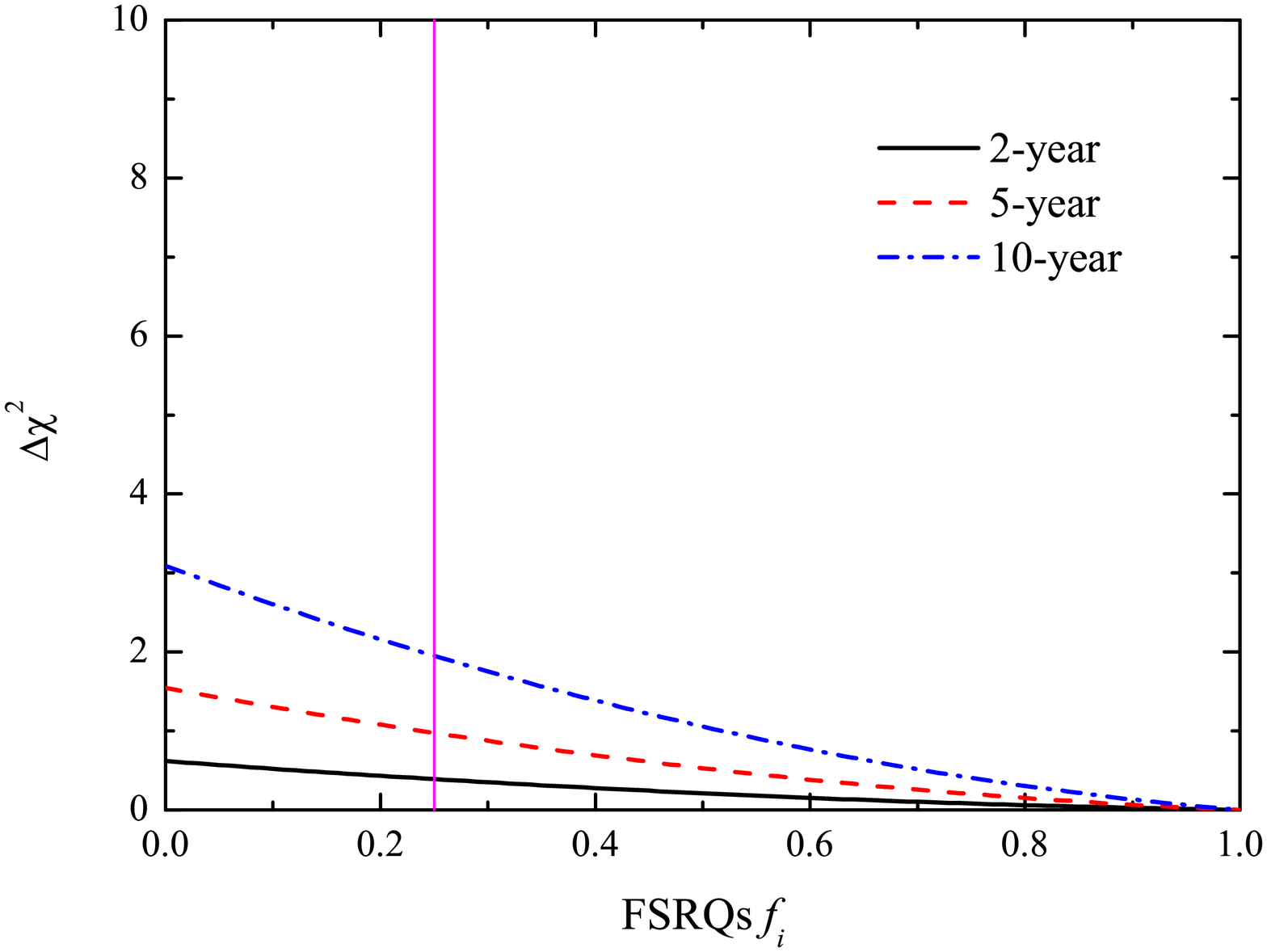, angle=0, width=0.33 \textwidth}
\epsfig{file=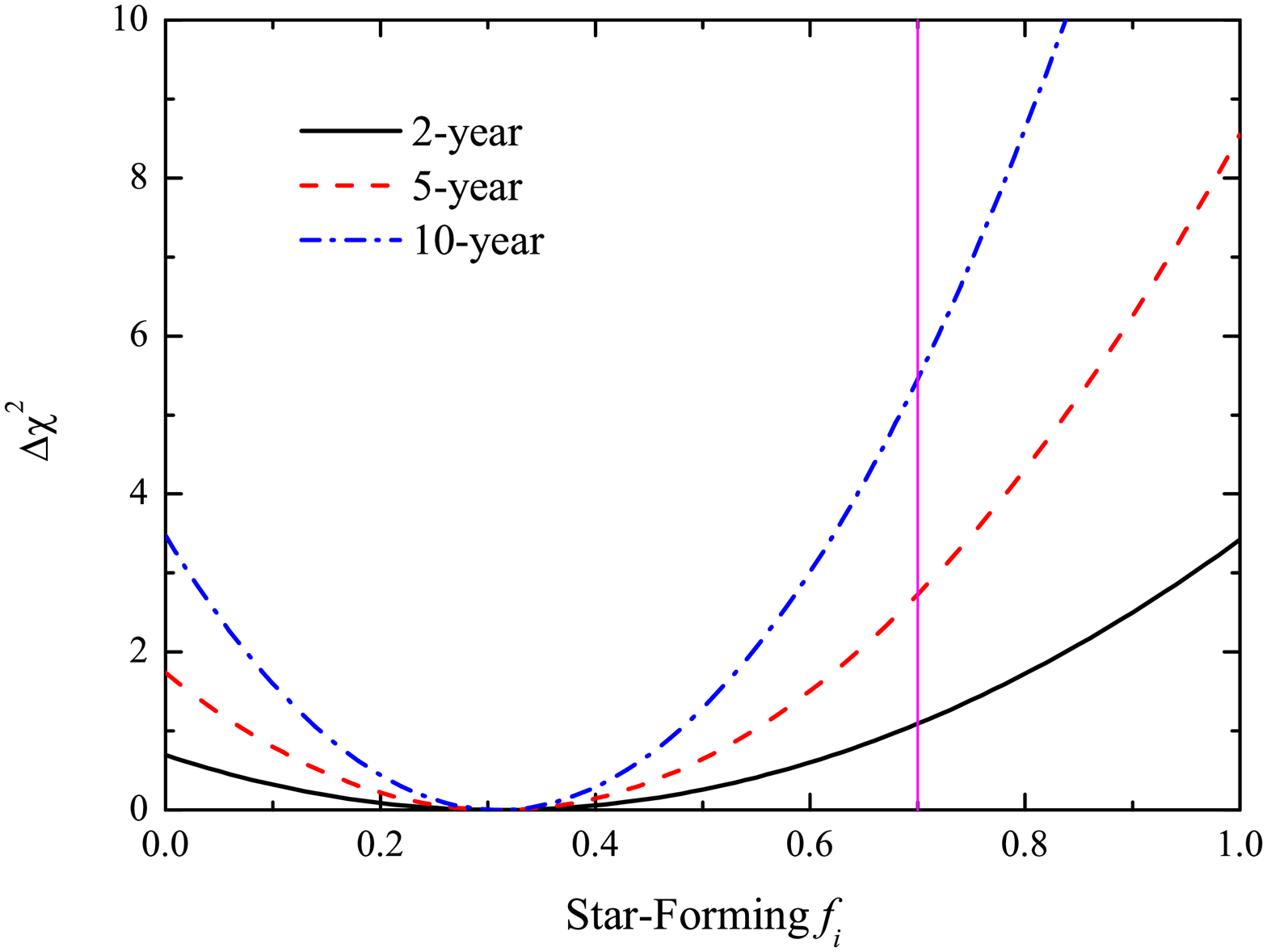, angle=0, width=0.33 \textwidth}

\caption{Limits on the fraction of three possible EGB sources from
the current 21-month Fermi-LAT data (black, continuous) and the
future 5-year (red, dashed), 10-year (blue, dot-dashed) Fermi-LAT
observations. The vertical lines denote the maximum values of
fractions from the current theoretical
constraints.\label{fig:fraction}}
\end{figure*}

As a second test we have computed the $\chi^2$ varying the fraction
of objects that contribute to the EGB, $f_{\rm j}$, in the interval
[0,1]. In this case we do not use the values listed in
Table~\ref{tab:tab1} but let the contribution of each source type
vary between 0\% to 100\%. As before, we do not explore the case of
a mixed contribution from different types of sources. The
sensitivity to $f_{\rm j}$ comes, in this case, entirely from the
linear dependence of the normalization of ACFs and CCFs on $f_{\rm
j}$ itself. We note that varying $f_{\rm j}$ would affect the
prediction for the $\gamma$-ray flux as a function of redshift. We
decided to neglect this effect given the large uncertainties in the
modeling of the sources. The results are displayed in
Fig.~\ref{fig:fraction}.  The plots show the $\chi^2$ as a function
of the EGB fraction contributed by the source $f_{\rm j}$.  The
three panels refer to the three models explored: FSRQs, BLLacs and
star-forming galaxies and, within each panel, the three curves refer
to different observation times: 21 months (continuous, black), 5
years (dashed, red) and 10 years (dot-dashed, blue). The vertical
lines are draw in correspondence of the reference values of $f_{\rm
j}$ listed in Table~\ref{tab:tab1}.  The minima of the $\chi^2$ do
not coincide with the fiducial values for $f_{\rm j}$. However, with
the current 21-month data the discrepancy is hardly significant
(barely above 1-$\sigma$ in the BLLacs model). We note that if one
removes the physical constraint  $f_{\rm j} \le1$ then for the FSRQs
the minimum of the $\chi^2$ would be found at $f_{\rm j} > 1$. The
fact that FSRQs favor large values of $f_{\rm j}$ is not surprising.
It simply reflects the fact that their $\gamma$-ray flux is
preferentially produced at large redshifts (see
Fig.~\ref{fig:zdist1}) and, as a consequence, that all  their
angular spectra plotted in Section~\ref{sec:results} have smaller
amplitudes that in the BLLacs and Star Forming galaxy cases. Notice,
however, that overall the sensitivity to the FSRQs component is very
weak and it is not expected to provide significant constraints even
after 10 years of data taking. The constraints will improve
significantly with the future Fermi-LAT maps: in the case of BLLacs
and star-forming galaxies, after 10 years of data taking one would
be able to reject the hypothesis of 100\% contribution to the EGB at
the  $\agt 3 \sigma$ confidence level (but only to $\sim 1.5 \sigma$
for the FSRQs).

\section{Discussion}
\label{sec:discussions}

The results presented in Section~\ref{sec:results} show that with
the current uncertainties it is not possible to discriminate among
the different models of source populations, considered in this
paper, contributing to the EGB emission. The measured ACF and CCFs
are generally consistent with zero and, in particular, the lack of a
measurable ISW effect prevents us from inferring the presence of a
cosmological constant. The $\chi^2$ test presented in
Section~\ref{sec:analysis}, however, demonstrates that the situation
will improve with the duration of the Fermi-LAT mission that, thanks
to the sheer number of the collected photons, will allow to reduce
the Poisson noise that contribute to the total uncertainties.

However, it is worth pointing out that, at present, the main factor
limiting the correlation analyses is not photon counts statistics
but the imperfect knowledge of the diffuse Galactic background which
prevents us from using the data at very low Galactic latitude and,
especially, in the lower energy band. Clues that  on some scales the
current Galactic diffuse emission is not perfectly modeled are found
when we compute the ACF and when we cross-correlate the 2MASS galaxy
and the QSO-SDSS catalogs with the residual Fermi-LAT maps obtained
with the lowest energy threshold $E>1$ GeV. In those cases we did
find a weak correlation signal at small angular separation $\theta
\le 2^{\circ}$ that, however, is very sensitive to the cleaning
procedure and that disappears when a progressively larger strip
across the Galactic plane is excluded from the analysis. A more
convincing correlation signal is detected at the same separations
and energy threshold when we cross-correlate the Fermi-LAT maps with
the distribution of the SDSS LRGs. This signal is remarkably robust
to masking and cleaning procedures. However this signal is above
theoretical expectations and is only detected at low energy.  {If
genuine, a positive cross correlation signal indicates that sources
contributing to the EGB are traced by LRGs and prompts some further
speculations. The fact that at $E\sim$ 1GeV the measured signal is
larger than expected may indicates that, at this energy, the bias of
the EGB sources ($b_\gamma$ in Eq.\ref{eq:autocorr}) is larger than
expected. In addition, the fact that the cross-correlation signal
drops for $E>3$ GeV may suggests that EGB is not contributed by a
single type of objects. Instead, the EGB would be produced by
different types of sources and their relative contributions depend
on energy. In this particular case, the decrease at $E=3$ GeV could
indicate that the relative contribution of FSRQs, whose
cross-correlation with LRGs is expected to be zero, increase with
energy. } Another possibility is that the $\gamma$-rays are
physically associated to LRGs just below the Fermi-LAT detection
threshold.

The model of Galactic diffuse emission is expected to improve in the
near term, when future dedicated analyses, based on a better photon
statistics, will be available. As a result it will be possible to
confirm or disprove these correlation signals and the validity of
the above speculations. Moreover, improving the Galactic model will
significantly increase photon statistics by extending the
correlation analysis to lower energy bands.

The actual reduction of the error bars will be more severe than what
expected from simple  statistics since a larger sky coverage will
allow to sample new structures, effectively reducing the Cosmic
Variance. Future cross-correlation studies will also benefit from
planned galaxy redshift surveys like
EUCLID~\citep{2009arXiv0912.0914L}, JDEM~\citep{2010arXiv1008.4936G}
and Big Boss~\citep{2009arXiv0904.0468S} that will be able to trace
the large scale structure over  a large fraction of the sky probing
the crucial epoch in which the current accelerated expansion of the
Universe has presumably started.

The main aim of the $\chi^2$ tests presented in
Section~\ref{sec:analysis} was to illustrate the limitations of the
correlation analysis of the current datasets and to show that future
datasets will probably allow to marginally detect the EGB
correlation signal and the ISW effect. Another  way to improve the
statistical significance of the future detections is to optimize the
tools to compare model with data and to consider additional objects'
catalogs in the cross-correlation analysis. Exploring the potential
of the various statistical tools that could be employed in a
cross-correlation analysis is beyond the scope of this paper.
However, even  with the paradigmatic case of the $\chi^2$
statistics, significant improvement can be obtained by restricting
the comparison to some particular range of angular scales, energy
range or to some particular catalogs of objects. From a more general
perspective, the most rewarding way of extracting astrophysical and
cosmological information from the diffuse EGB is to perform a
combined likelihood analysis combining the cross-correlation
analyses presented in this work with the one-point statistics,
angular auto-correlation and spectral information, as suggested by
\citet{2009PhRvD..80h3504D}.

We stress that the EGB models used in this paper can and will
improve significantly in the near future. Indeed, the number of
resolved extragalactic sources is bound to increase with time. On
one hand, this will allow us to resolve an increasing fraction of
the diffuse EGB. On the other hand, the next generation catalogs of
extragalactic source will allow to probe more reliably the faint end
of the log$N$-log$S$ of blazars and will provide better and better
estimates of their contribution to the diffuse EGB. Population
studies with higher statistics will also likely result in the
detection of the expected break in the luminosity function, allowing
to drop the assumption of a minimum $L_{\rm MIN}$. With strong
constraints on the blazars' contribution, one could afford exploring
the case of mixed contribution from different types of sources,
especially if the analysis can be performed over an extended range
of energies, since the contribution from star-forming galaxy is
expected to increase below 1 GeV. In this context one may also
include the possible $\gamma$-ray photons from DM annihilation in
extragalactic halos, whose contribution to the EGB has been modeled
by several authors (e.g.
\citet{2002PhRvD..66l3502U,2005PhRvL..94q1303A,2009arXiv0908.0195P}).

Finally, it is worth pointing out that investigating the ISW effect
using $\gamma$-ray data (even if on its own it does not provide
tight constraints on the cosmological parameters) is of considerable
importance since it would represent a consistency check for the
$\Lambda$CDM model obtained from a new, independent dataset. Current
ISW estimates that rely on different tracers for large scale
structure are sometimes in tension with the amplitude value for a
$\Lambda$CDM model (e.g. \citet{2008PhRvD..78d3519H}). The new
analysis performed in this paper can contribute to quantify to what
extent such measurements are reliable and identify possible
systematic effects.

\section{Summary and conclusion}
\label{sec:conclusions}

In this work we have extracted the maps of the diffuse EGB from the
21-month Fermi-LAT data by masking  $\gamma$-ray point sources from
the  Fermi-LAT data and by subtracting the various available models
of the Galactic diffuse signal.

These residual sky maps have been used to compute the angular
two-point auto-correlation function of the diffuse signal. To
minimize possible systematic effects that may affect the correlation
analysis, we have thoroughly checked for the presence of spurious
signals due to an imperfectly subtraction of the Galactic
foreground. In practice, we have tested the robustness of all
measured ACFs and CCFs to Galactic foreground models, cleaning
procedures and masking strategies. In doing so we have optimized a
strategy with combined cleaning and masking procedures that allows
to reduce  systematic uncertainties. We note that, as expected, the
ACF is much more prone to systematic effects than the CCF which, by
contrast, is remarkably stable.

In addition to the auto-correlation analysis, complementary to the
recent determination of the Fermi-LAT angular power spectrum by
\citet{2010arXiv1012.0755G} and by \citet{2010arXiv1012.1206S}, we
have cross-correlated these EGB maps with the WMAP7 CMB map, in an
attempt to detect the ISW effect. Finally, to unveil the nature of
the unresolved sources that may contribute to the EGB we have
cross-correlated the Fermi-LAT EGB maps with the angular
distribution of different types of extragalactic objects in several
catalogs. More specifically we have considered the local population
of 2MASS galaxies, that should trace the EGB contribution from
star-forming galaxies, and the more distant population of LRGs, NVSS
radio galaxies and QSOs that may trace the population of blazars.

All the measured ACFs and CCFs have been compared with theoretical
predictions in which the mean EGB signal and its angular correlation
properties have been modeled assuming only one type of contributing
sources: star-forming galaxies, BLLacs or FSRQs.  The results of
these comparisons allow, in principle, to constrain the level of
contribution to the EGB of the different sources and, mainly through
the ISW signal, the value of $\Omega_{\Lambda}$.

The main results of our analysis can be summarized as follows:
\begin{itemize}
\item

The measured ACF of the EGB above 3 GeV is consistent with zero at
all angular separations. This result is in agreement, within the
large error bars, with theoretical predictions since the
auto-correlation signal is expected to be very small in all EGB
models explored in this work. In other words, the auto-correlation
function does not seem to be the best statistical tool to reveal the
nature of the sources that contribute to the EGB. In the $E>1$ GeV
band we detect a $\sim 2\sigma$ positive correlation signal at
$\theta \le 2^{\circ}$ which, however, disappears when we increase
the size of the Galactic mask. Considering the sensitivity of the
ACF analysis on the Galactic diffuse model, especially at low
energy, we conclude that this signal is a spurious feature.

\item

The measured ISW signal in the energy band  is also consistent with
zero at all energies. The expected amplitude, however, is
significantly larger than that of the auto-correlation signal,
suggesting that future data with better photon statistics and more
accurate subtraction of the Galactic contribution could allow one to
detect the ISW signature.

\item

The CCFs measured considering various objects catalogs are generally
consistent with zero with a couple of exceptions: SDSS QSOs and
2MASS galaxies show a positive cross-correlation signal with the
$E>1$ GeV Fermi-LAT photons for $\theta \le 2^{\circ}$. However,
neither signal is robust to the cleaning procedure and the Galactic
mask applied. A more intriguing  correlation signal is found at
these same separations and energy when we correlate the Fermi-LAT
maps with the SDSS LRGs. Unlike the previous cases, this signal,
which is detected at $2 \sigma$, is remarkably robust to the model
adopted for the Galactic diffuse signal, to the procedure to clean
out spurious residuals and to the size and shape of the mask applied
to exclude the Galactic plane and the Bubble/Loop-I regions. The
analysis of future Fermi-LAT data will clarify the reliability and
the nature of all these features.

\item

A simple $\chi^2$ test performed using all the measured correlation
functions confirms, in a more quantitative way, that 21-month
Fermi-LAT maps allow to neither investigate the nature of the EGB
nor constrain the value of the cosmological constant. However, this
analysis shows that 10-year Fermi-LAT data would allow to constrain
the contribution of star-forming galaxies of BLLacs to the EGB with
a  confidence level of $\agt 3 \sigma$ and to confirm the presence
of a cosmological constant with a statistical significance of $\sim
2$ $\sigma$.

\item

These estimates are very conservative since they are based on a
simple extrapolation of the 21-month data assuming pure Poisson
errors. The Galactic diffuse model, however, is likely expected to
improve in the near future, while, new Fermi-LAT data will allow to
better constrain the contribution to the EGB of some some
extragalactic objects like the FSRQs and the BLLacs. As a result, we
will be able to extend the CCF analysis to lower energies,  further
improving the photon statistics, and providing more secure priors
for the $\chi^2$ analysis.

\end{itemize}

\section*{acknowledgments}
AC, EB and MF thank the Institute of Theoretical Physics, University
of Zurich and the Center for Cosmology and Astro-Particle Physics of
The Ohio State University for the kind hospitality. Their visits
were partially supported by NASA through the Fermi GI Program grant
number NNX09AT74G. MF thanks the support of the Spanish MICINN's
Consolider-Ingenio 2010 Programme under grant MultiDark
CSD2009-00064. MV is supported by INFN-PD51, ASI/AAE contract, a
PRIN MIUR, a PRIN INAF and the FP7 ERC starting grant "cosmoIGM".

We would like to thank M.~Ajello and M.~Ackermann for carefully
reviewing the manuscript and providing comments and suggestions
which have contributed to  sensibly improve the analysis. In
particular, we thank M.~Ajello for kindly providing the Montecarlo
simulation of the population of Blazars which we have used to
validate the analysis.

The \textit{Fermi} LAT Collaboration acknowledges generous ongoing
support from a number of agencies and institutes that have supported
both the development and the operation of the LAT as well as
scientific data analysis. These include the National Aeronautics and
Space Administration and the Department of Energy in the United
States, the Commissariat \`a l'Energie Atomique and the Centre
National de la Recherche Scientifique / Institut National de
Physique Nucl\'eaire et de Physique des Particules in France, the
Agenzia Spaziale Italiana and the Istituto Nazionale di Fisica
Nucleare in Italy, the Ministry of Education, Culture, Sports,
Science and Technology (MEXT), High Energy Accelerator Research
Organization (KEK) and Japan Aerospace Exploration Agency (JAXA) in
Japan, and the K.~A.~Wallenberg Foundation, the Swedish Research
Council and the Swedish National Space Board in Sweden.

Additional support for science analysis during the operations phase
is gratefully acknowledged from the Istituto Nazionale di
Astrofisica in Italy and the Centre National d'\'Etudes Spatiales in
France.

\bibliographystyle{mn2e}
\bibliography{version7.4}

\end{document}